\documentclass{article}

\usepackage{a4wide}
\usepackage{amssymb,amsmath,graphicx}

\usepackage{graphicx}
\usepackage{url}
\usepackage{hyperref}

\usepackage{xpatch}

\newcounter{mybibstartvalue}
\xpatchcmd{\thebibliography}{%
  \usecounter{enumiv}%
}{%
  \usecounter{enumiv}%
  \setcounter{enumiv}{\value{mybibstartvalue}}%
}{}{}

\usepackage{layouts}

\author{Krzysztof R. Apt$^1$ and Tony Hoare$^2$ \\[2mm] (editors)}
\title{Edsger W.~Dijkstra: a Commemoration}
\date{}
\begin{document}

\maketitle

\noindent
{\small $^1$ CWI, Amsterdam, The Netherlands and
MIMUW, University of Warsaw, Poland}

\noindent {\small $^2$ Department of Computer Science and Technology,
  University of Cambridge and Microsoft Research Ltd, \\
\hspace*{2.5mm}Cambridge, UK}

\begin{abstract}
  This article is a multiauthored portrait of Edsger Wybe Dijkstra that
  consists of testimonials written by several friends, colleagues, and
  students of his. It provides unique insights into his personality,
  working style and habits, and his influence on other computer
  scientists, as a researcher, teacher, and mentor.
\end{abstract}

\tableofcontents


\newpage

\phantomsection
\addcontentsline{toc}{section}{Preface}



\centerline{\bf Preface}
\bigskip\noindent
Edsger Dijkstra was perhaps the best known, and certainly the most
discussed, computer scientist of the seventies and eighties.

We both knew Dijkstra ---though each of us in different ways--- and we
both were aware that his influence on computer science was not limited
to his pioneering software projects and research articles.  He
interacted with his colleagues by way of numerous discussions,
extensive letter correspondence, and hundreds of so-called EWD reports
that he used to send to a select group of researchers.  His renowned
Tuesday Afternoon Club seminars, first in Eindhoven and later in
Austin, instilled in others his uniquely systematic way of approaching
research problems and developing solutions.  His courses at the
University of Texas in Austin were unlike any other, both in the
choice of topics and in the meticulous way they were delivered.

We felt that these aspects of Edsger's influence on the field might
become forgotten and next to impossible to reconstruct. In fact, some
of his collaborators and PhD students had died and his peers
are over eighty years old. Accordingly, we began the task of
documenting his impact on the life and work of his students,
colleagues, and scientific friends, and learning more about his
interactions with them.

We wrote to several researchers asking for a two-page contribution to
a collective article, mentioning that ``we are particularly interested
in evidence of Edsger's personal qualities, including kindness,
intellectual honesty, ideals, research standards, and
idiosyncrasies''.  It is telling that more than eighteen years after
his death so many computer scientists responded positively to our
request.  The result is this collection of over twenty uniquely
personal testimonials that convey a well-balanced appraisal of his
greatness and allow one to better appreciate Edsger as a researcher,
teacher, colleague, and friend.

We wish to express our heartfelt thanks to each contributor for their
willingness to share details of their relationship with Edsger and for
making it possible to produce a remarkable multi-authored portrait of
him in this way. We would also like to thank Rutger Dijkstra for
agreeing to include this article in the E.W. Dijkstra archive that he
curates.

This collection of tributes is designed for light reading.  If you have
only a short amount of time, read it in small parts.  If you print it
out, keep it as a bedside or a chairside book.  We have deliberately
avoided the traditional practice of numbered references, which have to
be looked up in a separate bibliography.
With a couple of exceptions the relevant titles will be quoted in full
in the main body of the text.
\bigskip

\noindent
K.R.A. and T.H. \\
Amsterdam, the Netherlands, and Cambridge, U.K. \\
March 2021
\bigskip

\noindent
PS. The second editor was the first to present his contribution of
personal reminiscences.  When the editors decided there would be more
space, he was the last to respond to this opportunity.  He added his
recollection of Edsger's scientific interactions.  The first editor
made the decision to present the combined material as the first rather
than the last tribute in the collection.

\hfill
\hypertarget{linkB}{\hyperlink{linkA}{$\hookleftarrow$}}

 \newpage

\phantomsection
\addcontentsline{toc}{section}{Tony Hoare}





\centerline{\bf Forty years with Edsger}
\centerline{Tony Hoare, Department of Computer Science and Technology,}
\centerline{University of Cambridge and Microsoft Research Ltd}
\centerline{29 March 2021}
\bigskip
\noindent
Edsger was a person of many idiosyncrasies: in his hates, his fears,
his clothes, his working environment, his work practices, and his
priority of values.  He had a strong aversion to cheese, both its
smell and its taste. With a similar aversion to cucumber, I
sympathised. He also had a distaste for broccoli, and other foods
containing dietary fibre. On taking guests to a standard Texas steak
joint, he was pleased to point out that chicken was included as a side
dish in the vegetable section of the menu.

He told me that he had once asked his audience whether they believed
that Computing was a Science, and why. A bold soul answered ‘Yes,
because computers exist’. His rejoinder was `So existence of broccoli
proves the existence of Broccoli Science?'. His choice here of
broccoli as an example was not fortuitous.

He had a strong fear of horses, because their behaviour is so
unpredictably agitated. I once took him for a walk from my home in
Oxford to a nearby meadow. To get there, we needed to take a narrower
section of the path, but clearly wide enough for walkers in opposite
directions to pass each other. On this section he saw approaching us a
man leading a horse. To me he said quietly that he thought we should
turn and go back. And so we did.

This fear explains the piquancy of his choice of a metaphor in the
discussion of the complexity of programs from his
\emph{Notes on Structured Programming}
(\href{https://www.cs.utexas.edu/users/EWD/ewd02xx/EWD249.PDF}{EWD249}):

\begin{quote}
  Apparently we are too much trained to disregard differences in
  scale, to treat them as ``gradual differences that are not
  essential'' $[\dots]$

  Let me give you two examples to rub this in. A one-year old child
  will crawl on all fours with a speed of, say, one mile per hour. But
  a speed of a thousand miles per hour is that of a supersonic
  jet. Considered as objects with moving ability the child and the jet
  are incomparable, for whatever one can do the other cannot and vice
  versa. Also: one can close one's eyes and imagine how it feels to be
  standing in an open place, a prairie or a sea shore, while far away
  a big, reinless horse is approaching at a gallop, one can ``see'' it
  approaching and passing. To do the same with a phalanx of a thousand
  of these big beasts is mentally impossible: your heart would miss a
  number of beats by pure panic, if you could!
\end{quote}

Edsger had a strong aversion to certain words and phrases and styles
of exposition, which were not uncommon in publications and lectures in
the field of Computing Science. He was particularly critical about
graphs of a function that had no measurements on their axes, and
diagrams that had no explanation of the meaning of their lines and
arrows. He hated words like `intuitive' and `natural', used as a
justification for a concept. He deplored the use of examples as a
motive (or even as a substitute) for proper definition. He regarded
all these faults as indicative of sloppy thinking.

On finding such a fault in reading a publication, he would take this
as a good reason to stop reading at this point.  In a lecture he would
often interrupt to advise the lecturer how and why to avoid them. In
his written trip reports, he often offended his hosts by helping them
with the same advice.

He gave long thought to the style and content of each paragraph of his
writing. He would not commit it to paper till he knew exactly the
complete text. He gave great consideration to the choice of his main
working tools and to his apparel. For writing his manuscripts, he
chose a Mont Blanc ink pen, which he replaced whenever the gold nib
was worn down.

In Austin he decided that the best headdress was a Texas cowboy hat;
the best for his collar was a Texas choker, which tightens around the
neck without a knot. For his trousers he chose a pair of shorts, and
for his feet a pair of sandals instead of knee-length jackboots
favoured by a true Texan. Instead of pockets he carried a leather
man-purse on a strap over his shoulder. It was an incongruous
combination, but one copied closely by several of his immediate
disciples.

I shared his distaste for operational semantics. An axiomatic
semantics can explain (maybe indirectly) the purpose of a concept and
how it can be properly used (e.g., a chair is for sitting in).  An
operational semantics would have to give instructions for building a
chair. In fact, he objected even to the term `Computer Science'. He
likened this to naming Astronomy as `Telescope Science', defining a
science by its instruments rather than by the area of its search for
truth.

Most highly, he valued simplicity. He used to be a heavy smoker.  He
gave up overnight after a visitor described the complicated algorithm
he had recently used to quit smoking. He thought `Surely it can’t be
that complicated' and proved it by quitting immediately, with no
subsequent relapse.

Edsger greatly valued brevity. He quoted from Confucius, the apology
for writing such a long letter, because he had no time to shorten
it. He frequently attributed to Confucius the precept that a picture
is worth a thousand words.  He just made a different claim, namely
that `A formula is worth a thousand pictures'.  Sometimes he followed
the contrapositive of Confucius' precept and drew highly illuminating
pictures; they were all worth more than a thousand words.

\begin{center}
    ***
\end{center}

My forty years with Edsger started in April 1961 in Brighton, England.
He was then just 30 years old, me nearly four years younger.  He was
lecturing on a course for programming in \textsc{Algol 60}, together
with Peter Landin and Peter Naur.  The latter had drafted and edited
the \textsc{Algol 60} Report, a complete and very clear definition of
the syntax and semantics of that language.  It was printed as an A5
booklet of 23 typewritten pages.  Yet it contained all the information
needed by an implementor of the language and by its user. And both of
them could understand it. I can vouch for this by personal experience.

When the class was set to solve exercises, I decided instead to write
an implementation of a recursive sorting algorithm that I had
discovered the year before, but had failed then to find an
implementation for.  I used recursion, which made it easy.  I showed
my solution to Peter Landin.  He was impressed, and beckoned to Peter
Naur to have a look.  But the third lecturer was standing some
distance away, and never saw it on that occasion.  He produced later
an elegant extension of it (smoothsort), which preserved the sequence
of elements with the same primary key.

Although I attended Edsger's lectures, we later both confessed that
neither of us could remember the other.  I attributed his lapse of
memory to the fact that I was only one student among many. He kindly
attributed my lapse to the difficulty of distinguishing between two
lecturers both wearing beards.

With me on the course was my future wife Jill, who was then my
colleague.  We both worked as programmers for a small British Computer
Manufacturer, a Division of Elliott Automation Ltd. We were
implementing an \textsc{Algol 60} subset to run on the next generation
of Elliott computers, which would be sixty times faster than their
current machine, but with the same architecture. I had documented my
design of the compiler in \textsc{Algol 60} itself.  It used recursion
to perform compilation on a single pass over the source code. This was
followed by a loader which resolved the destination of forward jumps.

\begin{center}
    ***
\end{center}

As a result of our experience in implementing \textsc{Algol 60}, we
were both selected as members of IFIP Working Group 2.1.  In September
1963 we both attended its second meeting in Delft.  This Group was
charged with the curation of the \textsc{Algol 60} language
---including corrections to the Report, the design of a subset of the
language, and of optional extensions to it. Finally, the design of a
language that would replace it.

The subset under discussion at the meeting had been designed by the
so-called ALCOR group of numerical analysts in the Group.  They
removed recursion from \textsc{Algol 60}, on the grounds of its
alleged inefficiency.  Edsger's answer was that recursion was a useful
programming tool, and every workman should be allowed to fall in love
with their tools.  I certainly had done so.  It is clear that Edsger
was a man after my own heart.

To my shame, when discussion turned to suggestions for extension of
the language, I reported a request of potential customers of my
Company's new machine.  It was to follow the example of \textsc{FORTRAN II},
and provide an option for omission of variable declarations together
with their types.

Kindly waiting until the next lunch break, Edsger approached me
together with Peter Naur.  They explained to me quietly just what a
bad idea it was. It would be a source of almost undebuggable errors
caused by the merest mis-spelling of an identifier. The only
protection against such errors was to build redundancy into the
programming language.  This would enable the compiler to detect source
program errors, and report them instead of running the program.

I made such checks the guiding principle for the implementation of the
Elliott \textsc{Algol} compiler. Because of its recursive structure, the
compiler detected all occurrences of all syntax errors and all type
errors and all scope errors.  And it produced code that detected all
run time errors, including subscript errors and numeric overflows.
These errors could make the program totally unpredictable from
inspection of the code, and therefore undebuggable.

This incident started me on a most productive direction for my career
of research. I spent ten years exploring the design of types for
variables and for structured data.  I published some of them
immediately in the
\href{https://archive.computerhistory.org/resources/text/algol/algol_bulletin/}
{Algol Bulletin}, and I finally assembled them under the title
\emph{Notes on Data Structuring} that eventually appeared as a chapter
in \cite{DDH72}. Edsger wrote the first chapter,
\emph{Notes on Structured Programming}.

The last meeting of WG2.1 that I attended was the one which made the
recommendation of \textsc{Algol 68} as a successor to \textsc{Algol
  60}.  It was also Edsger's last attendance.  We both abhorred the
complexity of the new language, and its manual of 140 journal pages.
Even the definition of its syntax used a cumbrous new notation, which
we found incomprehensible.  Others shared our misgivings, and under
the leadership of Edsger we wrote a Minority report (reproduced in
\href{https://www.cs.utexas.edu/users/EWD/ewd02xx/EWD252.PDF}{EWD252}),
discommending the language.  He drafted the first paragraph:

\begin{quote}
  We regard the current Report on Algorithmic Language \textsc{Algol
    68} as the fruit of an effort to apply a methodology for language
  definition to a newly designed programming language. We regard the
  effort as an experiment and professional honesty compels us to state
  that in our considered opinion we judge the experiment to be a
  failure in both respects.
\end{quote}

\begin{center}
    ***
\end{center}

The next ten years were our most intensive period of interaction of
our researches into the Theory of Programming.  The period began with
a seminar in 1971, held at the Queen's University, Belfast. The
lectures, together with subsequent discussions, were recorded in the
book \emph{Operating System Techniques} \cite{HP72}.  I gave a talk on
conditional critical regions.  They take the form:
\[
  \mbox{\textbf{with} $<$resource variables$>$ \textbf{when} $<$wake-up condition$>$ \textbf{do} $<$critical region$>$}
\]  
Other concurrent processes would have to discharge the responsibility
for waking up this process by making the wake-up condition true.

Edsger gave the next talk on
\emph{Hierarchical Ordering of Sequential Processes}.
At the end of it, he described his concept of the
`Secretary', acting as part of the kernel of a multi-programming
system. It consisted essentially of a set of conditional critical
regions sharing the same resource.  They were local to the secretary,
and scope checks would prevent any other process from accessing
them. These were called as procedures, possibly with parameters, by
the multi-programmed threads.  The secretary was slightly modified in
its scheduling capabilities by Per Brinch Hansen and me working
together. We renamed the resulting language feature as a Monitor.

Ideas of secretaries or monitors stimulated a new ten-year phase in my
research on concurrency, culminating in the publication in 1985 of a
textbook \emph{Communicating Sequential Processes}.

The following paragraphs examine the interactions of Edsger’s
discoveries with my design of the language of Communicating Sequential
Processes (CSP). They are interesting as a story of how two scientists
can start from different conceptual backgrounds (physics and
philosophy) and develop from them the same theory. It greatly
increases scientific confidence in the validity and wide applicability
of both the original theories.

The interaction will be described in greater technical
detail, as an extension of the conditional critical region of the
monitor concept, supplemented by the ideas expounded in Edsger’s
seminal 1975
\href{https://www.cs.utexas.edu/users/EWD/ewd04xx/EWD472.PDF}{EWD472}
\emph{Guarded commands, non-determinacy and formal derivation of programs}.
Edsger initially wrote all his proofs of programs
informally, and was resistant to the stark formality of Hoare
Logic. Grocery, he called it.

But in this article, he embraced formality enthusiastically. It led to
his own brilliant calculus of weakest preconditions, soon after used
in his book \emph{A Discipline of Programming}. It also introduced the
language of guarded commands. In my 1984 paper entitled
\emph{Programs are Predicates}, I suggested that the language can be reduced to a
simple set of algebraic axioms, much more elegant than the proof rules
of Hoare Logic. The interaction of our ideas was indeed intense.

In transferring Edsger’s ideas to CSP, the conditional critical region
of a monitor plays the role of a single guarded command. The syntax
had no selection of required resources, and was abbreviated to
\[
\mbox{$<$wake-up condition$>$ $\to$ $<$critical region$>$}
\]

The guarded command set was a collection of guarded commands,
connected by a nondeterministic choice operator $\Box$.  This selects
just one of the alternatives of the set.  The proof rule for the whole
set requires that the wake-up condition proof to guarantee the
existence of at least one such choice; and each possible choice must
satisfy the precondition of the following critical region.

The implementor is generally expected to choose the command whose
guard is the first to became true as in the external choice of
CSP. This scheduling strategy typically halves the expected length of
wait, and reduces the standard deviation even further. The resulting
reduction of unpredictable latency is significant in solving one of
the major complaints, both of interactive users of computers and of
the designers of cyberphysical systems. I summarised these benefits in
the title of a later lecture entitled
\emph{Concurrent programs wait faster},
where the paradoxical phrase was suggested by Edsger.

Edsger’s conceptual breakthrough in 1975 was to realise that whenever
two guards were simultaneously true, the implementer could make an
arbitrary choice. This was called `demonic choice', because the proof
of a program rules out all the possibly malicious choices made by the
demon. The opposite kind of choice is called `angelic choice', which
is postulated by Automata Theorists to simplify their analysis of
algorithms. In logic-based languages like \textsc{Prolog},
angelic choice is a requirement on the implementor, who has to use a
potentially exhaustive tree search to find a solution of a set of
implications expressed in predicate logic.

Formal development was just the replacement of a generally
non-determinate abstract design by a more concrete design which is
more deterministic. The opposite direction of design is called
abstraction, which is essential to formal development. It is normally
implemented by method declarations and method calls, as explained in
my 1972 \emph{Proof of Correctness of Data Representations} paper.

In the late seventies my research interactions with Edsger became
personal and my meetings then intensified. It included our
participation as lecturers and our joint Directorship of the
Markt\-oberdorf Summer Schools, including joint editorships of the
proceedings.  Edsger was involved as a lecturer, director, and editor
of the proceedings from 1972 until 1998 when he retired from the
Directorship.  I also spent a sabbatical year, visiting him at the
University of Texas at Austin (1986-7).

\begin{center}
    ***
\end{center}

Jill and I last met Edsger in his house at Nuenen, in August
2002.  He had frequently expressed the view that it was the duty of
the dying to comfort those who will survive them.  Accordingly, he and
Ria invited their foreign friends to visit him for a few days. We were
joined at mealtimes by more local friends.  He was completely lucid,
but too tired to talk about technical matters.

A few weeks later I was lecturing at the Summer School in
Marktoberdorf. I had the sad duty of announcing the death of their
much-loved Director. I was most comforted by recollection of our long
friendship with a man who embodied the best scientific, educational,
and personal traditions of the past. Indeed, in his humility about his
own intellectual powers and in his interactive method of teaching, he
was reminiscent of the ancient Athenian philosopher Socrates.
Socrates also comforted his friends as he went to his early death at
the hands of the Athenian executioner (Plato, \emph{Apology}).

I was invited to give a brief presentation at his funeral.  I cut
short my stay in Marktoberdorf to attend it. My speech ended with an
assessment that
\begin{quote}
  $\dots$ [he had laid] the foundations that would establish computing
  as a rigorous scientific discipline; and in his research and in his
  teaching and in his writing, he would pursue perfection to the
  exclusion of all other concerns. From these commitments he never
  deviated, and that is how he has made to his chosen subject of study
  the greatest contribution that any one person could make in any one
  lifetime.
\end{quote}

\hfill
\hypertarget{linkB}{\hyperlink{linkA}{$\hookleftarrow$}}


 \newpage

\phantomsection
\addcontentsline{toc}{section}{Donald Knuth}

\smallskip
\centerline{Don Knuth, Stanford Computer Science Department}
\centerline{29 November 2020}
\bigskip\noindent
I suspect that the first two people in history whose brains were perfectly
adapted for computer science were Alan Turing (1912--1954) and
Edsger Dijkstra (1930--2002). Thus it was a great privilege for me to have
had many encounters with Edsger, beginning in the early 1960s. My purpose
in this note is to share a few of them that still remain fresh in mind
as I think about him many decades later.

We first became acquainted through correspondence about programming languages.
Jack Merner and I had written a somewhat provocative note called
``ALGOL 60 Confidential'' [{\sl Communications of the ACM\/ \bf4} (1961),
268--272]. Edsger remarked in a letter that he didn't think it was
``written under the Christmas tree,'' and he recommended a less
confrontational tone. I tried to keep his advice in mind when I wrote
``The remaining trouble spots in ALGOL 60'' some years later
[{\sl Communications of the ACM\/ \bf10} (1967), 611--618].

Both he and I were part-time consultants to Burroughs Corporation during
those days, and he crossed the Atlantic at least once to visit their engineers
in Pasadena, near my home. One of the many things we discussed at that time was
his ingenious idea for a hardware mechanism that would guide an operating
system's page replacement algorithm, by maintaining a sequence of
exponentially decaying bits that recorded recent activity. I don't know if
he ever published those thoughts.

My files contain an interesting letter that he wrote on 29 August 1967,
enclosing a preliminary copy of
\href{https://www.cs.utexas.edu/~EWD/ewd02xx/EWD209.PDF}{EWD209}:

\smallskip
{\narrower\noindent
I seem to have launched myself into an effort to develop thinking aids that could increase our programming ability. 
\dots\
I expect that in the time to come this effort will occupy my mind completely.
\dots\
(This EWD209 was actually born when I reconsidered the origin of my extreme
annoyance with Peter Zilahy Ingerman, who to my taste failed to do justice to
Peter Naur in one of his recent reviews. And I started thinking why PZI's
attitude seems so damned unhelpful. So his unfair review may have served a
purpose, after all!)
\par}
\smallskip
In August 1973 I had a chance to visit his home in the Netherlands. By
coincidence, two of my most significant mentors, Dick de Bruijn (mathematics)
and Edsger Dijkstra (computer science) both lived in Nuenen, a few blocks
from each other. Edsger and I spent a delightful hour playing four-hands piano
music on his magnificent B\"osendorfer piano. (Of course I let him take the
lead in setting a proper tempo, etc.)

A couple days later I decided to have some fun when I gave a talk at the
Technical University of Eindhoven. I discussed what has become known as
the ``Knuth--Morris--Pratt algorithm,'' writing it in English step-by-step
on the blackboard. When I got to step 4, I paused and pretended to be
at a loss; I said, ``Hmmm. Is it legal to use the words `{\bf go to}' in
this place?'' Edsger said, ``I saw it coming.''

My wife snapped a nice picture of Edsger and me when he stayed at our
Stanford home on 18 April 1975.

I think he had an encyclopedic knowledge of just about everything.
During dinner conversations we rarely discussed computing, and the topics
varied widely. I always learned something new from him, whatever the subject.

During January 1993 I stayed a few days with Ria and Edsger at their
Austin home. We drove to Pedernales Falls State Park, and I was pleased to see
that they both were genuinely in love with Texas.

\smallskip

\centerline{\includegraphics[width=1\textwidth]{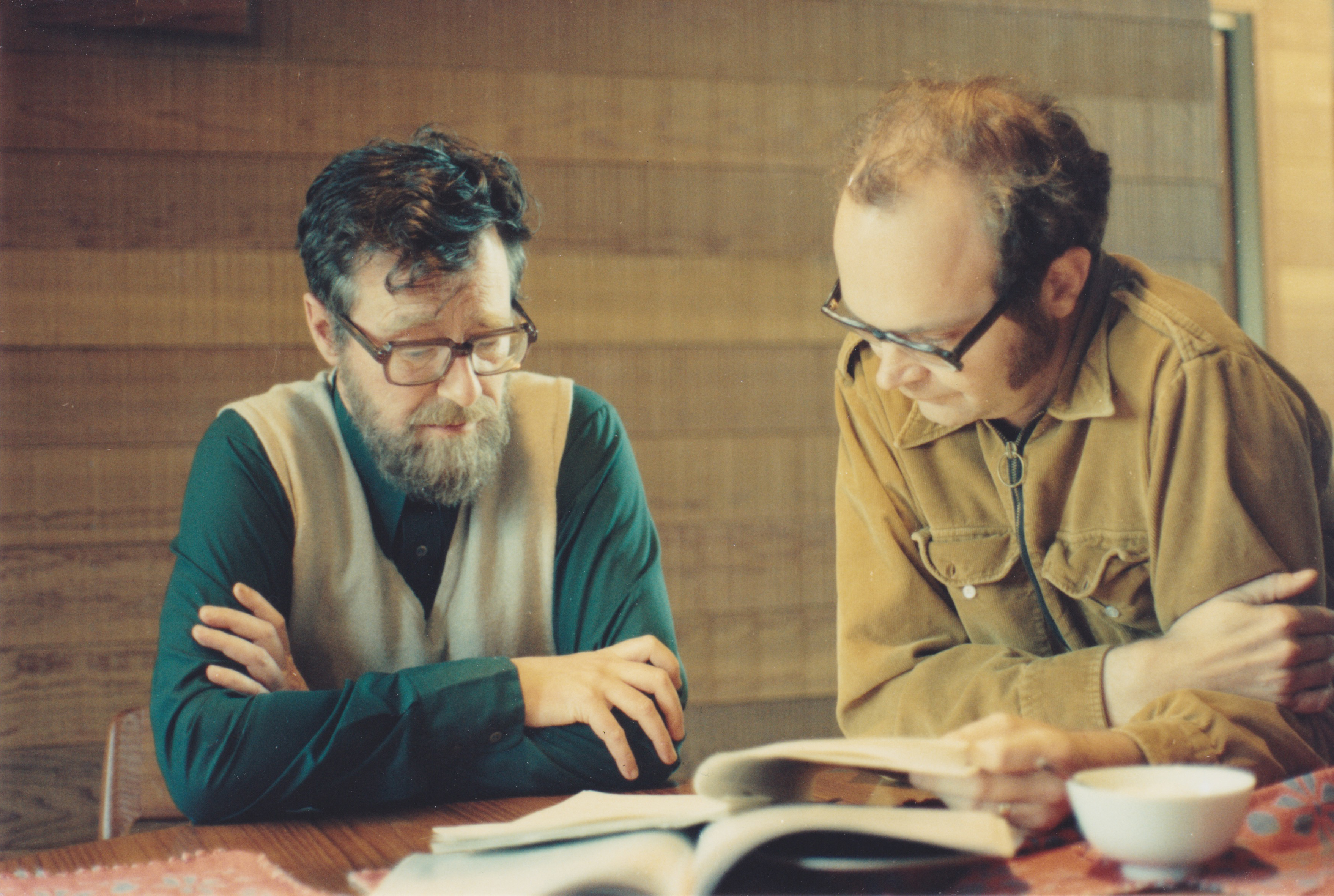}}

\hfill
\hypertarget{linkB}{\hyperlink{linkA}{$\hookleftarrow$}}

 \newpage

\phantomsection
\addcontentsline{toc}{section}{Christian Lengauer}


\centerline{Christian Lengauer, University of Passau}
\centerline{22 December 2020}
\bigskip\noindent
When I studied the book on structured programming as a mathematics student
at the Free University of Berlin in the mid-Seventies I could not have
fathomed that I would ever interact with Edsger W. Dijkstra---let alone
gain his friendship. But when I had joined UT Austin on an assistant
professorship in the Eighties, he joined two years later. He offered me a
membership at his Austin Tuesday Afternoon Club (ATAC) and I also had the
pleasure of experiencing the generous and sociable side of Edsger and Ria Dijkstra.

Some time after they had got settled, I was invited for dinner at the
Dijkstra home.  It was my first time with them alone. Like many
others, I was a bit in awe of Edsger.  Dinner was delicious and the
conversation was relaxed and turned to many topics, with pauses in
between. Edsger had the knack of starting to speak just when you were
ready to say something---a point of synchrony to which I had to get
used at first.

I had just returned from an extended trip to Europe and recounted, among
other things, innocently a flight from Saarbrücken to Berlin-Tempelhof in
a Fokker \mbox{F-27}, the popular
Dutch twin-propeller airplane from the Fifties. What
made the flight notable were the vibrations that I felt
especially since I sat right under the wing. I had to keep my coffee
cup from walking across the table. After the touchdown in Berlin, when
the engines had died down, a meek voice from the back of the plane
broke the silence: “You still alive (\emph{Lebste noch})?”

I thought that funny and entertaining, but Edsger remained silent and
Ria responded quietly: ``But it’s a good plane...''
Both were clearly taken aback, I supposed because the F-27 is a Dutch
product. Later that evening, Edsger told me about his past and the
beginnings of his career as a ``computational engineer'' in the Netherlands.

I departed one hour before
midnight. When I waited at the long traffic light downhill from their
home for green, I replayed in my mind the things that had been said---and
only then I realized that Edsger had quietly worked into the
conversation \dots\ that he himself had been involved---not
in the development, I suppose---but in the coding of \emph{the wing stability
  equations for the Fokker F-27}! I felt touched by his tact and frankness towards
me, a junior colleague whom he barely knew.

Some time later, Edsger called me at home and announced that he and Ria would
like to come over. I cannot remember any other senior colleague of mine ever
having invited himself this way. My latent alarm whether there was a serious
reason was unfounded. They just wanted to spend time chatting and sharing a
drink. I had acquired recordings of the Vienna New Year concerts of the early
Eighties and played Strauss waltzes when they arrived. After an hour I switched
to Haydn. Edsger inhaled cigarette smoke with pleasure and said quietly:
``Ah, the music has improved.''

He and I enjoyed exchanging classical music via self-recorded cassette
tapes. He always wrote out the contents of his tape most carefully on
the cassette cover, with recording date and dedication to me.  This
way I learned that he was not fond of waltzes or organ music---but not
before I had brought him an organ record from the Passau cathedral
with, at the time, the largest church organ world-wide. He had
accepted the gift without complaint.  Edsger did love his
B\"osendorfer piano which occupied a prominent place in his living
room.  I remember his delight when it arrived, still completely in
tune, after its long journey from Nuenen to Austin.

Edsger and Ria showed me their interest and affection not only in
Austin.  Maybe there was a mutual kinship between us as Europeans in
Texas.  When I attended PARLE'89 in Eindhoven, they insisted that I
should stay at their house in Nuenen for the week.  At another visit
two years later, Edsger, wearing his ``Don't mess with Texas.''
T-shirt, took me on a tour on the backseat of his tandem bike (picture
taken by Ria). A truly unique experience!

Professionally, Edsger's impact on me is best summarized by his ATAC
Rule~0: ``Don't make a mess of it.'' It made me strive for simplicity
in notation and modelling throughout my working life and take
unpleasant complexity as an indication of a possible lack of
comprehension.  In my previous work at the University of Toronto,
under the supervision of Rick Hehner, I had become familiar with the
weakest precondition calculus and the view of programs as mathematical
objects, and so I felt very much at home in Edsger's world.

Having had the privilege of being around Edsger continually, it was
easy to avoid the traps of his dislikes and I adopted some but not all
of his preferences. My research group propagated his quantification
notation and proof format in Passau. But I did not start numbering
from zero and, after a failed attempt, I decided against writing with
a fountain pen. He never took issue with this, except in a letter that
he posted seven weeks before his death: ``I hope you will overcome
your resistance and learn how to fill a pen without soiling your
fingers, or otherwise you are denying yourself one of the joys of
life.'' These were his last written words to me.

Edsger W. Dijkstra had strong convictions and lived by them without compromises. It was a
true pleasure and an eye-opening privilege to have been part of his life.

\smallskip

\centerline{\includegraphics[width=0.8\textwidth]{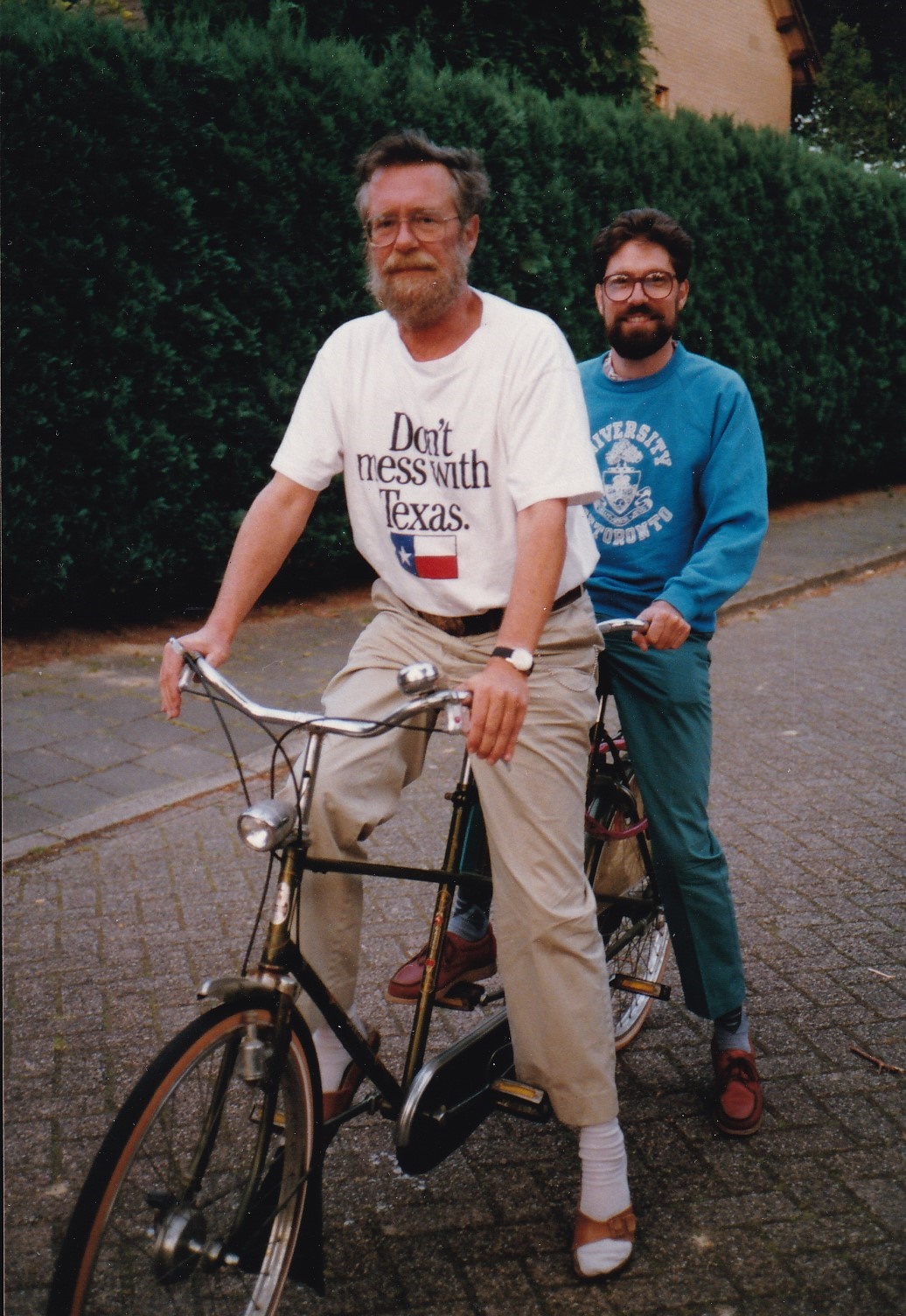}}
\hfill
\hypertarget{linkB}{\hyperlink{linkA}{$\hookleftarrow$}}

 \newpage

\phantomsection
\addcontentsline{toc}{section}{K. Mani Chandy}

\centerline{K. Mani Chandy, California Institute of Technology}
\centerline{18 January 2021}
\bigskip\noindent
I was the Chair of the Computer Sciences Department at the University
of Texas at Austin when the university made an offer to Edsger to join
the department. Many had misgivings. Everybody agreed that
Edsger was a great computer scientist, a genius. Everybody also agreed
that he was among the most outspoken, blunt, opinionated scientists on
the planet. Moreover, Edsger was proud to belong to that group. Edsger
was particularly skeptical about Artificial Intelligence and was known
to voice that skepticism. The CS department at UT was a broad
church. And broad churches survive on tolerance while Edsger was
notoriously intolerant. Nevertheless, we made Edsger an offer and I
think it was one of the best decisions that the department
made. The decision was not without controversy then and remains
controversial even now.

Edsger influenced many people in the department. The Austin Tuesday Afternoon Club
(ATAC) meetings were a delight. His book with Carel Scholten,
\emph{Predicate Calculus and Program Semantics}, taught us
the beauty of calculational proofs. Edsger required
that we read papers aloud at the ATAC. This seemed to me to be an
old-fashioned tradition from days when scholars actually “read papers”
as opposed to modern presentations of colourful slides and videos. I
learned, however, that reading aloud exposed redundancies, unnecessary
adjectives, and shoddy writing. Flashy presentations hid all of
that. I still read my papers aloud to myself.

As the chairman of Edsger’s department, I sometimes found him
painfully blunt. Not for him, the gentle reprimand. He also had a
wicked sense of humor. For example:
\begin{itemize}
\item
He asked my wife and me on hearing that she had a PhD in psychology:
“How do you two manage to stay married?” Edsger must have felt that psychologist
-- CS relationships were unstable.
\item
His words of encouragement when he found out that I graduated from
MIT: “I am happy that you seem to be working at overcoming your poor
education!”  
\item
While I was giving a chalkboard talk, he announced: “I’ve never met an
American computer scientist who could give a talk without drawing two
boxes and joining them by a line!”
\end{itemize}

Edsger and Ria used to drop in unannounced at our house for a chat and
a drink, a nice habit common in India but less so in the US. My
three-year old son, getting ready for his bath and bedtime story, came 
downstairs in the nude, saw Edsger and said --– much to Edsger's delight
–-- “Oh no! Not you again!” 

I often had Edsger over for dinner with students, and he was always
entertaining and pleasant, belying his reputation of intellectual
haughtiness.  Edsger was extremely kind, a kindness he hid in his
gruff exterior. For example, he was friendly to my parents apparently
enjoying their conversation, though my father and mother couldn’t have
been more different than Edsger and Ria.

Jayadev Misra and I had lunches with him regularly at the Santa Rita
room at UT, and these were wonderful, happy times. He was often funny
with Oscar Wilde-like quips. He also educated the two of us, over
these lunches, about programming: what it meant to him and what it
ought to mean to us. Edsger was very helpful to us, reading
our papers, making helpful suggestions: encouraging
while remaining a Dutch Uncle. When I write papers, even now, I still
see Edsger over my shoulder going “Tsk! Tsk!”

Here's a little example of how Edsger influenced us. We used
implication, $P \Rightarrow Q$ as a predicate at times and as a
boolean at other times, expecting our audience to understand the
meaning by context. And the audience almost always got the intended
meaning. Edsger, however, thought that was sloppy. He and Scholten
introduced the square bracket to identify booleans, as in $[P
\Rightarrow Q]$. Just a little notational thing, but an important
one. I found that being less sloppy not only helped my readers
understand what I had written, but most importantly, helped me from
confusing myself.

Edsger was \textit{the} major influence on Jayadev Misra and me as we
wrote our book on concurrent computing. Edsger's impact shows in our
calculational proofs and the emphasis, throughout the book, on
formalism and logic. If, instead, we had used anthropomorphisms ---
which are quicker to write, and perhaps more readable --- we would
have finished sooner and perhaps had a wider audience. But, we have
never regretted the Dijkstra-effect and remain forever grateful.
\hfill
\hypertarget{linkB}{\hyperlink{linkA}{$\hookleftarrow$}}

 \newpage

\phantomsection
\addcontentsline{toc}{section}{Eric C.R. Hehner}

\smallskip
\centerline{Eric C.R. Hehner, Department of Computer Science, University of Toronto}
\centerline{2021 January 19}
\bigskip\noindent
I have had a career-long interest in formal methods of program design,
and I still teach a course with that title.  My introduction to the
subject was the book \emph{a Discipline of Programming} by Edsger
W. Dijkstra in 1976.  I wrote my first paper on the subject that same
year.  So I was eager to meet Edsger when he came to Toronto for the
IFIP Congress in 1977.  He had read my paper and agreed to meet me.
He introduced his just completed PhD student Martin Rem and me to each
other.  Martin and I connected both personally and in our research
interests, freeing Edsger from baby-sitting duties.

Following the IFIP Congress, Edsger went to an IFIP Working Group 2.3
meeting in Niagara-on-the-Lake, where I was the newly appointed
secretary of the group.  My paper, which was critical of some aspects
of Edsger's work, was a topic of discussion.  Edsger was a very
important person, and I was nobody, and that provoked several people
to defend Edsger.  But Edsger found merit in my criticisms.  He was
more willing to consider how his own work could benefit from the
criticisms than some of the others in the room.

After the working group meeting, Edsger had a couple of days to kill
before his flight back to the Netherlands, so he came to my house.  My
mother was also visiting just then.  Back then, my mother and Edsger
were smokers, and I didn't allow smoking in the house, so they went
out on the front porch together.  My mother happened to mention that
she was not fond of existential proofs with no instantiating example
(witness).  Well, after that, Edsger and my mother talked only to each
other; I might as well have been dead.

Four years later I had the opportunity to return the visit, staying in
Nuenen.  Since their son Rutger was away, I was given his bed.  As we
were heading out for a ride on the tandem bicycle, Edsger's wife Ria
noticed a hole in the back of my pants.  Edsger tried to shush her,
but too late.  He had noticed it too, but he didn't know if I had a
replacement pair.  He reasoned that I would be better not knowing
there was a hole than knowing and unable to do anything about it.
Fortunately, I did have another pair.  But it stuck in my mind as an
example of Edsger's kind consideration, quite opposite to his stated
philosophy of ``telling truths that hurt''.

After Edsger moved to Austin, he held the Year of Programming in
1986-1987.  A variety of formal methods researchers spent all or part
of the year in Austin, and I was fortunate to be included.  This put
me in almost daily contact with Edsger for several months.  One of the
benefits for me was sitting in on Edsger's undergraduate course on
Mathematical Methodology.  I sat with Wim Hesselink because we were
somewhat out of our age group.  Each class Edsger arrived with a
problem to be solved.  The lesson was how the formal expression of the
problem guides the solution, or in his words, ``let the symbols do the
work''.  What I learned in that class permeates my formal methods
course today.

Another benefit to me of the Year of Programming was participation in
the Tuesday Afternoon Club.  I had attended one or two of these in
Eindhoven, and in Austin I attended several more.  Each week we read a
paper, chosen by Edsger; someone read aloud so we were synchronized
and the pace was slow.  Anyone could ask a question or make a comment
or criticism at any time, and the criticism could be at any level,
from syntactic (how well is the idea expressed) to semantic (how valid
is the idea) to judgemental (how important is the idea).  I was amazed
at how productive and effective this format is.  On my return to
Toronto, I instituted a copy of the Tuesday Afternoon Club (but not on
Tuesday afternoon).

From 1977 to 2001, about every 9 months, I had the pleasure of talking
and dining with Edsger at IFIP Working Group 2.3 meetings.  At one
such meeting one evening, Edsger saw that I was stuck listening to a
known pompous bore.  Edsger came across the room, apologized for
butting in, and said that I was urgently needed to settle an argument
in his group.  He rescued me.

Towards the end of his life, Edsger became concerned about his legacy.
One day, walking to lunch, Edsger said to me ``In a hundred years, the
only thing left of all my work will be the shortest path algorithm.''.
He may have expressed the same thought to others because J Moore said,
In Memoriam, ``Without a doubt, a hundred years from now every computer
scientist will study Dijkstra's ideas, including 
\begin{itemize}
\item the mathematical basis of program construction, 

\item operating systems as synchronized sequential processes, and
  
\item the disciplined control of nondeterminacy,
\end{itemize}
to name but three.''.

In my last letter to Edsger, I said ``All of us have taught literally
thousands of students, who take away thoughts that originated in your
head.  And your influence is not just technical.  Your way of working,
and your ethics, have become mine as well as I am able to emulate
them.  What I want to say is: thank you.  With great admiration and
affection, Rick''.  His last letter to me, handwritten of course, dated
Nuenen, Friday 19 July 2002, just 18 days before he died, said ``I
thank you for your friendship.''.  I treasure that letter.

Here is a picture taken in Austin in 1990 at Edsger's 60th birthday
party.  He asked me to sit next to him at the head table.  (To my
right is Elaine Gries.)  Note Edsger's bolo tie.
\smallskip

\centerline{\includegraphics[width=1.0\textwidth]{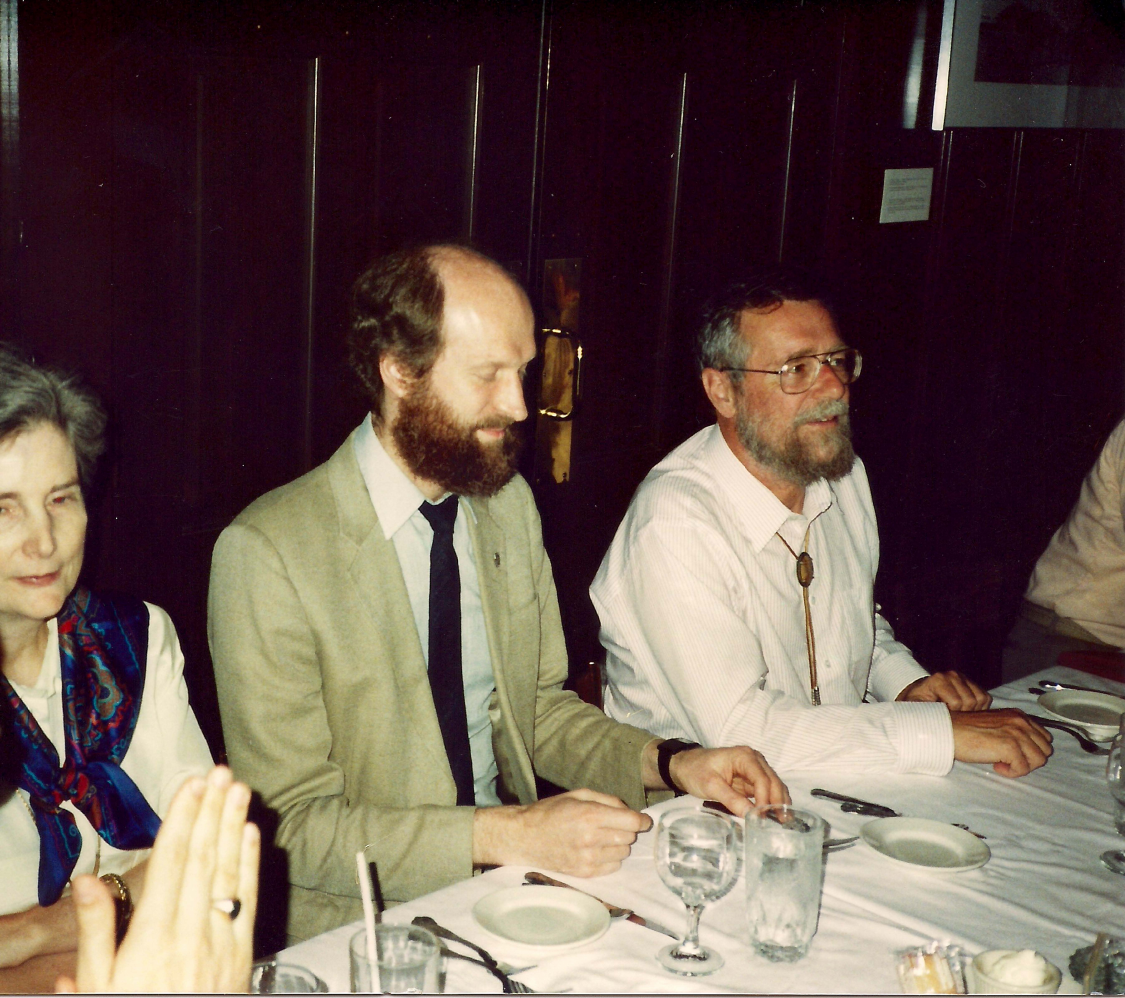}}
\hfill
\hypertarget{linkB}{\hyperlink{linkA}{$\hookleftarrow$}}

 \newpage

\phantomsection
\addcontentsline{toc}{section}{Mark Scheevel}


\centerline{Mark Scheevel, Vericast}
\centerline{31 January 2021}
\bigskip\noindent
After receiving degrees from Rice University in 1978 and 1979, I moved
to Austin where Burroughs Corporation had established a research
center to study approaches to parallel processing, among other
things. We were studying functional programming languages as a means
to achieve significant parallelism without requiring explicit
synchronization and coordination by a programmer.

Although it may not be widely remembered today, Edsger was a Burroughs
Corporation research fellow from 1973 to 1984. He also spent quite a
bit of time in Austin, eventually accepting a position at the
University of Texas in 1984. This meant that he was a frequent visitor
to our research center, and he spent many hours educating us and
critiquing our work.  As was common at the time, I came to computer
science via electrical engineering (in fact, Rice didn't formally
establish a department of computer science until 1984). That meant
that I did a substantial amount of assembly language programming in
addition to programming in higher-level languages, and I developed a
tendency to regard programs as something one hammered into shape over
time. Edsger introduced me to the idea that programs could be treated
as mathematical entities, with properties that could be subjected to
mathematical analyses, and that there were ways to prove assertions
about the properties of a program.

Those are powerful ideas, and they had a profound effect on the way I
approached problems. In my career, I have worked on compilers,
interpreters, query languages, and query optimizers, and in each of
these areas these ideas have been invaluable. At an abstract level,
each of these is about translating a source language to a target
language. This entails creating a semantic model for those languages,
identifying how the elements of that model are represented in those
languages, and devising rules for translating source language patterns
into target language patterns while preserving the semantics of the
source language patterns. Now, I do not mean to suggest that I created
formal models for these languages, or that I built formal proofs of
the correctness of the translation processes. For one thing, I am not
a proficient enough mathematician to do so. But understanding that
abstract plan helped me formulate better solutions, and I believe that
those solutions were better than they would have been had I not been
introduced to those ideas.

I have had many influences in my career ---and I am indebted to
several of my Burroughs colleagues who helped a young engineer find
his footing--- but I was truly fortunate to have the opportunity to
interact with Edsger during my early, formative years. He made me a
better engineer.
\hfill
\hypertarget{linkB}{\hyperlink{linkA}{$\hookleftarrow$}}

 \newpage

\phantomsection
\addcontentsline{toc}{section}{Krzysztof R. Apt}


\centerline{Krzysztof R. Apt, CWI Amsterdam and University of Warsaw}
\centerline{7 February 2021}
\bigskip\noindent
I first met Edsger Dijkstra in the spring of 1975.
Initially, our contact was very sporadic ---for instance, I met him
for the second time more than two years later--- but over the
years our contact grew more frequent. Eventually we became
colleagues in the same Department ---at the University of Austin---
and ended up exchanging letters throughout the nineties, followed by faxes,
and, when Edsger yielded in the last two years of his life to the
Internet, by emails.

When I contacted Edsger in 1975 I was looking for an academic job in
computer science, but knew nothing about the subject; in fact, I had
learned what a flowchart was just a couple of weeks earlier.  I was 
25 years old at the time, with a PhD in mathematical logic from Warsaw,
while Edsger was almost twenty years senior, with the Turing Award in
his pocket.  Even though I arrived late to our appointment, he allowed a
full hour for our meeting.

I left with two book recommendations: Niklaus Wirth's \emph{Systematic
  Programming: an Introduction} and \emph{Structured Programming}
edited by him, Tony Hoare, and Ole-Johan Dahl, an advice to learn
\textsc{Pascal}, and with a copy of his recent manuscript
\href{https://www.cs.utexas.edu/users/EWD/ewd04xx/EWD472.PDF}{EWD472}
titled ``Guarded commands, non-determinacy and formal derivation of
programs''. Little did I know it would play an important role in my
scientific career.

On the train from Eindhoven back to Nijmegen, I browsed through the
short article and was disappointed to notice it contained only some
trivial observations and a couple of obvious theorems, with no proofs,
all concerning some proposal of a simple programming language.  My
opinions were confirmed a couple of years later, when during my
employment at the Mathematisch Centrum in Amsterdam, Jaco de Bakker
explained to me how Dijkstra's guarded commands language could be
trivially modelled in the so-called relational calculus.

In contrast, Dijkstra's book suggestions were immediately
invaluable. Together with his 1976 book \emph{A Discipline of
  Programming} they made me aware early on that some computer programs can
be viewed as masterpieces, just like poems, in which not a single word
should be changed.

My first moment of doubt about my assessment of Edsger's article
arose while visiting Gordon Plotkin in Edinburgh in the early
eighties. He showed me his thorough mathematical analysis of various
semantics of Dijkstra's language.  So it was a language proposal worth
studying after all.  This, and a simple program example from
Dijkstra's 1976 book which showed that the assumption of fairness
implies unbounded nondeterminism motivated our work on the latter.

Dijkstra's example also triggered my interest in fairness.  The
outcome was an article with Ernst-R\"{u}diger Olderog about
correctness of guarded command programs under the fairness assumption.
I cannot now imagine a more elegant framework to explain and study
fairness.  We followed this work by studying correctness of parallel
programs under the fairness assumption, and their natural translation
to guarded commands programs provided us with key insights.

Tony Hoare quickly understood the potential of the guarded commands
language and extended it in 1978 to an elegant proposal of a
distributed programming language called Communicating Sequential
Processes (CSP).  A year later Nissim Francez, Willem-Paul de
Roever and myself came up with a Hoare-style proof system to prove
correctness of a large class of CSP programs.
Some time later I realized that a much easier proof system would do
for simpler CSP programs.  When working with Ernst-R\"{u}diger on our
book on program verification, which eventually appeared in 1991, we
realized that the easiest way to justify this simpler proof system was
to first provide a natural transformation of these CSP programs into
---what else--- the guarded commands language.

It was only then that I understood how vitally important simple and
elegant concepts are in computer science.  It had taken me more than
fifteen years to learn this simple lesson.

Returning to my contacts with Edsger, a major change occurred in 1984,
when I was invited to be a lecturer at the Marktoberdorf Summer
School, where the enclosed photo of the two of us was taken.  During
that summer I realised that Edsger visibly enjoyed his many roles as a
lecturer, a session chairman, and a co-organiser, but it was striking
to see how awkward and stiff he became when sharing a lunch or dinner
table with others. Small talk was not his cup of tea.

In 1987 I joined the Department of Computer Science at the University
of Austin, where Edsger had held a prestigious position since 1984.
During the interview I found Edsger to be remarkably informal and
accessible.  For instance, when I mentioned that I sought a specific
CD record of Thelonious Monk, he nodded approvingly and immediately
offered to drive me to the best music store in town that evening.

While in Austin I took Edsger's course on writing mathematical proofs.
Homework consisted of writing down proofs of elementary, but
non-trivial, mathematical results. Edsger returned them a week later
with detailed comments, upon which he presented his own proof,
impeccably explained, without any notes, using a piece of chalk and a
blackboard.  Edsger's elegant course substantially changed my
understanding of what a properly written proof should look like.

My times in Austin also allowed me to experience Edsger in yet another
role. Not only was he a departmental colleague and a dedicated
teacher, but he was also somebody who, together with his wife Ria,
enjoyed inviting guests to their home, and occasionally just liked
to pop in with her in the evening for an informal chat.

After I returned to Amsterdam in 1990 we kept in touch by occasional
letters. In 2000 I flew to Austin to participate in Edsger's
retirement. Whilst there, Edsger accepted my invitation to give a
lecture at CWI, the successor of Mathematisch Centrum, his first
employer. He delivered it in October 2000. (Indirectly it led to a
fascinating half hour long TV programme about him available at
\url{https://www.youtube.com/watch?v=-Uae9_pgZzE}.)  A year later,
upon learning that he was incurably ill, Edsger and Ria returned for good to
their home in Nuenen. I met them there for a lunch in July 2002.  It
was a very sad visit: Edsger told me calmly that he would die soon.
As a farewell he handed me his last EWD,
\href{https://www.cs.utexas.edu/users/EWD/ewd13xx/EWD1318.PDF}{EWD1318}.
Just like 27 years earlier, I read it in the train from
Eindhoven. Edsger passed away a month later.

It is difficult for me to summarise Edsger's influence on my research,
as absorbing his ideas is a never ending story. For example, in 2018 I
published a joint article with Ehsan Shoja which related Edsger's
self-stabilisation notion to strategic games. Only then did I understand
what a marvellous concept it is. And I keep discovering new for me
EWDs that amaze me.

I feel privileged to have met Edsger and to have experienced first-hand
his unmatched brilliance, originality, and integrity.

\smallskip

\centerline{\includegraphics[width=1\textwidth]{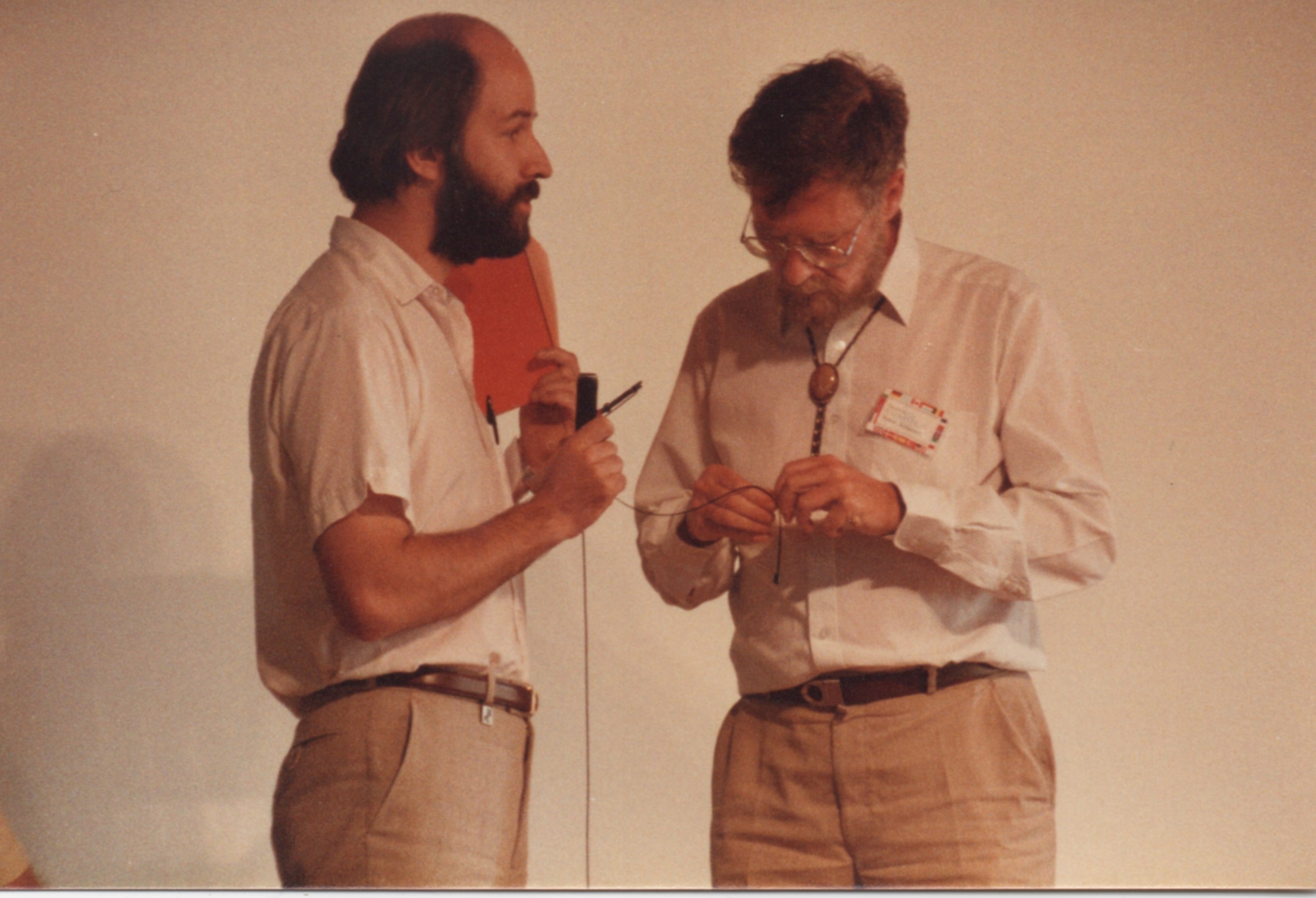}}
\hfill
\hypertarget{linkB}{\hyperlink{linkA}{$\hookleftarrow$}}

 \newpage

\phantomsection
\addcontentsline{toc}{section}{Niklaus Wirth}


\centerline{Niklaus Wirth}
\centerline{8 February 2021}
\bigskip\noindent
Edsger was a remarkable person, an icon in Computing Science. His
early achievements were the first full compiler for the language
\textsc{Algol 60}, handling recursive calls of procedures as well as
the controversial name parameter, and the operating system THE for the
Dutch computer X-8, said to have been provably correct.

As of 1948 Edsger studied physics at the University of Leiden. In 1951
he followed a three-week course on programming in Cambridge, which led
the following year to his employment as a programmer at the
Mathematisch Centrum in Amsterdam. Since 1962 he was professor at the
Technische Hogeschool Eindhoven, a position he combined as of 1973
with a consulting job at the Burroughs Corporation that he carried out
from a smallest possible, one person, office created at his
home. During his employment at the University he got entangled in
struggles with his anti-computing mathematics colleagues. This
obviously could not last forever. He was offered a professorship at
the University of Texas at Austin that he accepted to many people's
surprise, who knew about his America-critical attitude.

He was a stern man, tall and self-assured, bespectacled and with beard
and moustache. In early times he was always armed with a
pipe. Occasionally he used a long “reading pipe”, until once, deep in
thought, he bumped it into a door and hurt himself in the throat. Then
he switched to cigarettes. Another of his characteristic personal
tools was his fountain pen, notably a Montblanc. He wrote all his
memos and papers with this pen in a perfect hand-writing. And he wrote
hundreds of his ``EWD'' memos distributed to a select, private mailing
list.

I met Edsger Dijkstra for the first time at the big IFIP Congress in
Munich in 1962, where he presented a prominent and impressive paper
titled \emph{Some meditations on Advanced Programming}. He reflected
on the rapidly increasing power of computers, resulting in ever
increasing demands on programmers, and the inadequacy of the human
mind and the tools available. The paper was an early call for methods
to prove a program’s correctness, that it performed what
specifications promised, in short, for a more rigorous, mathematical
treatment of the subject. His high ambition was nothing less than
promoting programming from a craft to a science. Later on he coined
the term \emph{mathematical engineering} for it. Edsger was genuinely
worried about mankind’s inability to cope with the challenges of
computing technology and the road it was taking. When in 1964 IBM
announced its System/360 computer family concept, he claimed that this
caused him several sleepless nights. He became a member of the new
Working Group of IFIP, intended to define a successor to the language
\textsc{Algol 60}. His was a prominent voice in the 40 member club,
but his concern was programming discipline rather than language.

In September 1965 he wrote the memorable monograph \emph{Cooperating
  Sequential Processes}
(\href{https://www.cs.utexas.edu/users/EWD/ewd01xx/EWD123.PDF}{EWD123})
proposing the \emph{semaphore} as the essential, indispensable element
for synchronizing parallel processes. The $P(s)$ operator was to
inspect a (binary) semaphore $s$ and set it in an \emph{atomic}
operation, and $V(s)$ would reset it. The paper’s main innovation,
however, was the \emph{critical section}, pieces of the respective
processes to be executed under mutual exclusion, to be guarded by
semaphores.

The other influential monograph was \emph{Notes on Structured
  Programming}
(\href{https://www.cs.utexas.edu/users/EWD/ewd02xx/EWD249.PDF}{EWD249})
in 1969. It demanded a careful, well-structured approach to program
design with clearly nested statements for conditional and repeated
execution. Preceding it had come his warning against unstructured
code, epitomized by the harmful goto statement creating “Spaghetti
code”. This monograph had a profound influence on the design of the
language \textsc{Pascal}, and even more on \textsc{Modula} and
\textsc{Oberon}, as well as later indirectly on \textsc{Java} and
\textsc{C\#}. It was also the precondition for proving analytically
programs to be correct. The monograph had also been the ignition for
the movement called \emph{Software Engineering}. This was, however,
not quite in his spirit, and he called it disdainfully “Programming
when you can’t”.

After Hoare had published his work on axiomatic proofs of program
correctness, Dijkstra proposed his predicate transformers, essentially
replacing Hoare’s triples (precondition $p$, statement $s$, postcondition
$q$) by the function $p = \emph{WP}(s,q)$.

Many of his illuminating slogans were circulating, perhaps the most
profound one being “Testing can at best show the presence of errors,
but never their absence”. In short: Exhaustive testing is
impossible. Thus he underscored his demand for formal proofs. As an
illustration, he presented the following minimal example (which is
elegant but hardly convincing, because it would require trying two
cases only):

\begin{quote}
Consider two bottles. One of them contains a treasure. Bottle $A$
carries the inscription: “The treasure is not in here”, the other the
inscription $B$: “Only one of the inscriptions is true”. Where lies the
treasure? Without “trying and testing”, we deduce

($\neg A = B) = B$    ($=$ is associative)

$\neg A = (B = B)$

$\neg A = \mathbf{T}$   (the treasure is in bottle $A$).
\end{quote}

Another of his recommendations was: “Do not give industry what it
wants, but rather what it needs”. There was much truth in it, but
nevertheless it smelled of academic arrogance. And lastly:
“Universities should be leaders, and not followers”. Both statements
were strongly directed toward the choice of programming languages and
tools.

One of Edsger’s most peculiar idiosyncrasies (after 1970) was his vow
never to use a computer. Even after receiving a computer as a gift he
would obstinately place it on the lowest shelf of some cabinet and let
it rest there. He did not want to become bedraggled with the nasty
details of a genuine gadget. Diddling with a real computer would only
distract from the essentials of design. This explains a lot. In fact,
he had always shown a certain disdain for engineers. He considered
them as tinkerers. They showed an abominable lack of mathematical
knowledge, and thus were forced to resort to trying and
testing. Hence, so his thinking went, one is better off and free of
unpleasant frustrations when avoiding their questionable concoctions.

Edsger had contributed significantly to a well-founded, rigorous
approach to programming, in establishing it as an engineering
science. I therefore awaited eagerly the appearance of a textbook, a
fundamental guide to programming from his pen. But he had lost his
interest in this endeavor, and instead concentrated exclusively on
mathematical treatments and theories. They often originated in the
Tuesday Afternoon Club, a select group of colleagues and students to
discuss deeper problems and analysis of programs. His memos
increasingly resembled a collection of rigorous solutions of
(mathematical) puzzles rather than remedies for “the programming world
out there”. This was definitely an unfortunate loss, as his “voice in
the desert” gradually faded away. It will not be replaceable.

But fond memories remain. In my case these are conferences and
visits. Edsger and his wife Ria were always most gracious hosts,
socially active within their admired circles. They also learned to
like the country and Texas. They acquired a VW camper bus and
travelled Texas-style distances in their famous “Touring Machine”. They
started to love this way of outdoor life and thereby perfected their
transition from old European to American life style.

I enclose a photo of Tony Hoare, Edsger, and myself made near Einsiedeln in March 1971.

\smallskip

\centerline{\includegraphics[width=1\textwidth]{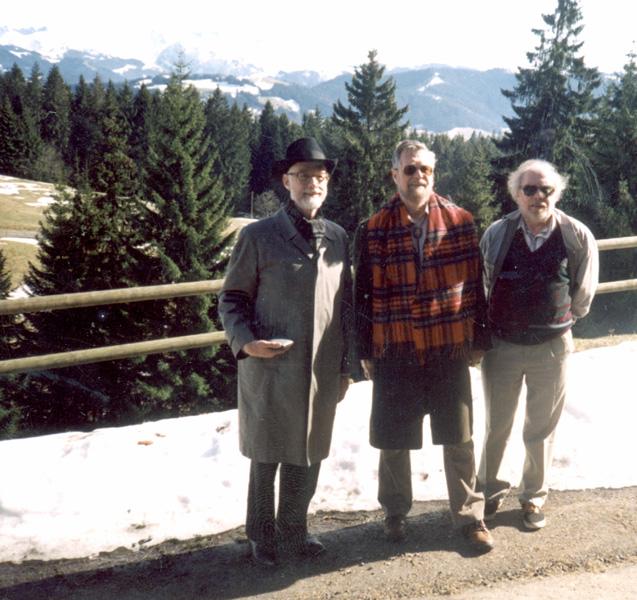}}
\hfill 
\hypertarget{linkB}{\hyperlink{linkA}{$\hookleftarrow$}}

 \newpage

\phantomsection
\addcontentsline{toc}{section}{Lex Bijlsma}
\centerline{Lex Bijlsma, Open University of the Netherlands}
\centerline{9 February 2021}
\bigskip
\noindent
First and foremost, I am a mathematician. Until I met Edsger, I never
had anything to do with computing. I did my Ph.D.\ thesis on the
theory of trans\-cendental numbers, and for the next five years I
continued to work at that subject. Doing so, I absorbed the prevailing
attitude among my fellow researchers, namely that a problem is solved
when only finitely many cases are left to check. Most theorems in this
area simply state that there exists a computable number $C$ such that
all integers greater than $C$ satisfy some hypothesis. This is
supposed to solve the problem, as the reader is invited to check the
numbers up to $C$ by hand. Never mind that, in the few cases where
someone actually took the trouble to compute $C$, it turned out to
exceed the number of particles in the universe: finite domains were
considered to be trivial on principle. Because of the finiteness of
machine memory, these mathematicians considered all of computing
science, with the possible exception of Turing machine theory, as too
simple for serious attention. (I once asked the dean of the Faculty of
Mathematics and Computer Science, a distinguished mathematician, how
long it would take for computing to be regarded as a full-fledged
independent discipline, no longer a poor relative of mathematics. His
answer: about 3000 years.)

This was the world in which I was living when, in February 1983,
Edsger Dijkstra, whom I knew by sight but had never spoken to, visited
my office to invite me to attend a session of the Tuesday Afternoon
Club, his famous weekly exercise in thinking together. At the time,
the group was working on the problem that became `Reducing control
traffic in a distributed implementation of mutual exclusion'
(\href{https://www.cs.utexas.edu/users/EWD/ewd08xx/EWD851.PDF}{EWD851}).
I suspect my being invited was due to curiosity as to how a
working mathematician would look at such problems. It is not
inconceivable that my reactions contributed to the way mathematicians
are described in `On a cultural gap' (
\href{https://www.cs.utexas.edu/users/EWD/ewd09xx/EWD913.PDF}{EWD913}).
Nevertheless, I became
intrigued, not so much by the subject itself as by the disciplined
thought habits cultivated in this circle and by the astounding
efficacy with which these could tackle quite challenging problems. To
make a long story short, one year later I had metamorphosed to a
sufficient extent to be invited to a permanent membership of the
Tuesday Afternoon Club. I attended every session for fifteen years and
I cannot think of any influence that shaped my mind more, except
possibly learning to read (taught by my grandfather, years before
school).

A famous quote of EWD is the following: `I mean, if 10 years from now,
when you are doing something quick and dirty, you suddenly visualize
that I am looking over your shoulders and say to yourself ``Dijkstra
would not have liked this'', well, that would be enough immortality
for me' 
(\href{https://www.cs.utexas.edu/users/EWD/ewd12xx/EWD1213.PDF}{EWD1213}).
I can testify that this actually works. I cannot
introduce a notation without wondering whether all the subscripts and
parentheses cannot be eliminated and whether the size and the
symmetries of an infix operator symbol accurately reflect the
properties of the abstraction it signifies. And it took me several
minutes to decide if the introduction of identifier $C$ in the first
paragraph was justified.
\hfill
\hypertarget{linkB}{\hyperlink{linkA}{$\hookleftarrow$}}
 \newpage

\phantomsection
\addcontentsline{toc}{section}{Manfred Broy}


\centerline{Manfred Broy, Technical University of Munich}
\centerline{15 February 2021}
\bigskip\noindent
The first time I met Edsger W. Dijkstra was in the year 1978 at the
Technical University of Munich where I was working as a young PhD
student and research assistant with Professor Fritz Bauer and
Professor Klaus Samelson. One day, Fritz Bauer informed me that Edsger
W. Dijkstra was about to visit him and since during the time of
Edsger’s visit, Fritz happened to be busy with another commitment for
one hour, he asked me to talk to Dijkstra during this time.

So, I had to chance, being a very young scientist, to talk to Edsger
W. Dijkstra, who was at the time perhaps the most prominent computing
scientist, for more than one hour. I still remember our conversation
very well. It was most remarkable that Dijkstra seemed obviously quite
interested in our interaction. After I told him what I was
scientifically working on, Edsger carefully discussed my topics of
research and answered my questions. This is just one example of Edsger
being very encouraging to young scientists.

Only a few months later I took part in the Marktoberdorf Summer School
where Edsger was one of the lecturers. All the participants were
fascinated by the particular style of Edsger’s lectures. He did not
bring in any prepared slides as all the other lecturers did, but wrote
everything by hand during his lecture. In his lectures, he developed
small but intricate programs in a very systematic and logically
precise way. He used to say that the programs he presented were like
\emph{little poems}. His lectures were very impressive and also highly
motivating.

In the summer school, there was a discussion period every day at the
end of the lecture hours. In this discussion, Edsger behaved very
special. Participants could ask questions. For each question, Edsger
gave a careful but sometimes a bit lengthy answer. Moreover, he wrote
little mysterious quotes on the blackboard. Often during the
discussion period, he was also asking questions to other lecturers. In
doing this, he was sometimes quite belligerent. But not only in the
discussions, also during the lectures of other lecturers he was
carefully paying attention, all the time sitting in the first row and
asking questions. Soon it became obvious that he did not like some
lecturers for their presentation style and the subject of their
lectures. Not only he criticised their way of lecturing, but also the
way they had written their material on the slides, questioning their
terminology, and sometimes even interrupting and disturbing their
lectures in some rude manner. In his striving for computing science
being a serious scientific discipline he did not hesitate to be very
explicit and even offending colleagues who were offering scientific
approaches he did not accept. This attitude is also reflected in a
number of his writings, including, for instance, his claim that our
discipline should rather be called “computing science”, and not
“computer science”.  I learned to appreciate his attitude as an
expression of demanding a precise terminology and as a plea for a
highly careful way of speaking, thinking, and acting at a scientific
level.

He also liked to take quite extreme positions. I remember a quite
absurd conversation between Edsger and Fritz Bauer on the eve of one
of the Marktoberdorf Summer Schools. At that time, Fritz Bauer
prepared and installed the exhibition on informatics at the Deutsche
Museum in Munich. Fritz was eager to show this exhibition to Edsger
and offered him a guided tour through the exhibition. Edsger said the
suggested visit would be difficult or even impossible for him because
he was too sensitive against visual impressions. Therefore, he rather
would not like to go for a visit. So Fritz offered Edsger that
he would guide him blindfolded all the way to the one piece of
exhibition ---an early Zuse computer--- that he wanted to show to
Edsger, in particular, and then Edsger could take off his blindfold
and just look at that piece. Edsger thought about this proposal for a
while and then he said that he cannot do it ---it would be too much
for him. What an absurd conversation.

Later, being the managing editor of Acta Informatica, I exchanged
quite some correspondence with Edsger who was one of the main
editors. In a number of cases, when I got submissions that were in
some sense borderline cases I used to ask Edsger for his referee’s
report and advice. And I must say, I always got the most impressive
and carefully handwritten reports by Edsger, very much to the
point. However, sometimes his reports were a bit of an insult to the
authors of the submitted papers. In these cases, I started to reword
and to retype Edsger’s reports a bit before I sent them out to the
authors to avoid that they felt insulted.
 
In the 80s, I spent a research stay in Austin. There, I met Edsger a
number of times. He invited me to the Tuesday afternoon club and I
agreed to take part in it. The club studied a paper on a
generalization of Edsger’s wp calculus. In this calculus, there is one
tricky point related to unbounded nondeterminism. In Edsger’s book, “a
discipline of programming”, Edsger decided to define the semantics of
the do loop by functional iteration instead of fixpoints ---actually,
he avoided unbounded nondeterminism in his book and usually did not
like operational reasoning. Only later, Edsger found out that a fixed
point definition would have been more appropriate. The text studied by
the Tuesday afternoon club showed a serious error in some definitions
about weakest preconditions due to the effect that it contained
nondeterministic constructions that were not continuous but only
monotonic. To my surprise, Edsger immediately took off the telephone
and called the young scientist who had written the text. This poor guy
was quite shocked to get a call by the great Edsger Dijkstra to tell
him that the Tuesday afternoon club found out a difficulty in his
text. Again, an example how strict Edsger was in his quest for a
flawless scientific discipline of computing.

In the end, I admire and appreciate Edsger Dijkstra for his seminal
contributions to our field, for his uncompromised scientific position,
for his consequent fight for what he considered to be scientifically
right, and for his unbroken striving for a scientific foundation for
our field ---fighting against large ignorant crowds found in our field
too often till today.
\hfill 
\hypertarget{linkB}{\hyperlink{linkA}{$\hookleftarrow$}}

 \newpage

\phantomsection
\addcontentsline{toc}{section}{David Gries}


\centerline{David Gries, Department of Computer Science, Cornell University}
\centerline{18 February 2021}
\bigskip\noindent
In summer 1972, Edsger and I taught week-long courses at Maryland at
the same time. I taught compiler construction; Edsger, operating
systems. We sat in each other's lectures. We discussed my text on
compiler construction, parts of which he took issue with. Our wives,
Elaine and Ria, were there, and we got along well. Elaine and I taught
them how to throw a frisbee, which had just become popular. Thus began
a friendship that lasted until his death.

Edsger ---and Tony Hoare--- profoundly influenced me, and I soon
switched my research from compilers to programming methodology and
related topics. I hesitate to think what a flop my career might have
been if I hadn't met these giants.

Edsger visited Cornell fairly often after that, and his daughter,
Femke, studied here for a year. To get a taste for what Edsger was
like, read his report of a 1980 trip to Cornell (see
\href{https://www.cs.utexas.edu/users/EWD/ewd07xx/EWD727.PDF}{EWD727}). In
a way that no other computing scientist could do, Edsger weaved a
journal of his trip with honest opinions of the places and people he
visited, a tongue-in-cheek list of ``language rules'', and his view of
the problems faced by CS in the US. After reading this EWD, you will
want to read more.

I didn't feel it then, but I now view the 1970's and early 1980's as a
magical time. The Marktoberdorf Summer Schools were high points. About
ten faculty would lecture to 100 PhD students from around the
world. Also, in the annual meetings of IFIP Working Group 2.3 on
Programming Methodology, people talked less about what they had done
and more about what they were doing. Edsger had a great deal to do
with how the Summer Schools and WG2.3 meetings were run. This quote
from 
\href{https://www.cs.utexas.edu/users/EWD/ewd07xx/EWD714.PDF}{EWD714}, ---a
must read--- shows you what Edsger thought of WG2.3 and introduces you
to his scientific mission.

\begin{quote}
"[We had] the shared recognition that $\dots$ only the most effective
application of mathematical method $\dots$  could be hoped to solve the
problems adequately. The charter was clear: searching for relevant
abstractions and separation of concerns that are specific to
programming and learning how to avoid the explosion of formulae when
dealing with stuff more complicated than mathematicians had ever
formally dealt with before. I found this charter sufficiently
challenging to devote my scientific life to it; at the same time this
charter was sufficiently unpopular ``in the real-world'' $\dots$ to justify
the protection of an IFIP Working Group $\dots$"
\end{quote}

In my opinion, three characteristics helped make Edsger \emph{the} dominating
figure in computing. First was his brilliant mind. Who else could
develop the shortest-path algorithm in their head in 20 minutes while
having coffee with their fiancé? But who else would then recognize
that ``brainpower is by far our scarcest resource''?  (From his Turing
Award speech, \emph{The Humble Programmer}.)

Edsger spent his career thinking about method, in programming and
mathematics. Tony Hoare provided tools for proving programs correct,
with his axiomatic basis for computer programming. But Edsger showed
us how to \emph{develop} programs using those tools. Edsger also spent his
scientific life contributing to mathematical method as no one else has
done. It was enlightening in a Marktoberdorf Summer School to see him
present calculational logic, which led to principles and strategies
for developing formal proofs. But I am much more awed reading his
EWDs, which are filled with developments of arguments and proofs,
discussions, developments of proofs and programs, always with an aim
of illuminating \emph{method}.

Second, Edsger had a passion for computing and its growing
importance. Here's what he said in his Turing Award speech about IBM
360 computers in the mid 1960s: ``$\dots$ the design embodied such
serious flaws that I felt that with a single stroke the progress of
computing science had been retarded by at least ten years; it was then
that I had the blackest week in the whole of my professional life.''
This was based on his knowledge of the problems of concurrency and
interrupts, for he had helped develop both a computer and its
operating system in the 1950s and 1960s.

Third, Edsger wrote and spoke with honesty and candor. Gerry Salton of
Cornell said about one of his trip reports, ``Dijkstra's right, but we
don't say such things.'' Edsger would have replied, ``If you don't say
such things, how can we hope to improve?''

Here's what happened to me at a Marktoberdorf Summer School in the
1970's. Writing an assignment like ``$x:= 5$'' on the board, I said ``$x$
equals 5.'' From the back of the room came a loud, booming, Dijkstra
voice, ``BECOMES''. I was startled, but I regained my composure and
said, ``Thanks. If I make that mistake again, tell me.'' I made it once
more during that lecture. You can envision what happened. I have never
made that mistake since then.

Edsger critiqued not the person but only what they said, and later one
could drink a beer and laugh as if nothing happened. Technical
differences and shortcomings should be treated this way.

Few people knew that Edsger and Ria were social animals. During our
1989-90 sabbatical year in Austin, we had dinner at each other's
houses, alternating weeks. The evenings would be spent talking,
discussing a CS issue, or reading a new chapter of the book ``A Logical
Approach to Discrete Math'' I was writing with Fred
Schneider. Sometimes, Edsger would play his piano, a Bösendorfer.

Also, Edsger and Ria would drop in unannounced on an evening. A knock
on the door, and in they came. The first time it was a shock, but we
soon looked forward to such visits. They visited several people in the
Austin CS Department in this manner.

There's so much more I could say about Edsger, but space does not
permit it. The best way to learn more about this giant is to read his
EWDs. My scientific life was spent working and playing with this
giant, and I am ever so grateful that we met.

The photo shows Manfred Broy, Edsger, and me during a skit one evening near the end of
the Marktoberdorf Summer School in 1998.
\smallskip

\centerline{\includegraphics[width=0.9\textwidth]{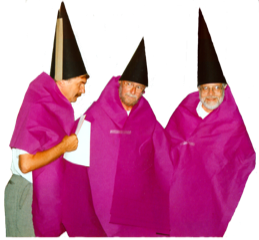}}
\hfill
\hypertarget{linkB}{\hyperlink{linkA}{$\hookleftarrow$}}


 \newpage

\phantomsection
\addcontentsline{toc}{section}{Ted Herman}


\centerline{Ted Herman, Department of Computer Science, University of Iowa, Iowa City}
\centerline{19 February 2021}
\bigskip\noindent
During the eighties I was a PhD student at the University of Texas at
Austin. My advisor, Mohamed Gouda, suggested I take a course taught by
Professor Dijkstra, which was my first exposure to programs as objects
of formal treatment. A while after finishing that course, I was
invited to attend the Tuesday Afternoon Club, which I'd not heard of
and about which I had no expectations. I remember being surprised
because my performance in the course was hardly stellar.  Aims and
activities of the club are set down in [0]. A subjective account of
the experience, from a beginner's vantage, follows.

Usually at the start of each meeting, paper copies of the material to
be read were distributed; one person, would then read aloud sentence
by sentence.  Attendees might suggest corrections, note their
confusion, and question choices of definitions: reading progress was
deliberate and not rushed. Generally the document under consideration
was authored by one of the members, if not Edsger. Of course a
blackboard was handy, should the need arise to work out some point. In
these slow readings it was easy to detect superficial flaws, a lack of
explanation, and missed opportunities to use better notation.  By the
conclusion of a meeting, there was deeper appreciation of the content
as well. Some meetings ended with homework for us students: finishing
a proof, working on a conjecture, or going over the remainder of an
unfinished paper.

Club meetings were intended inculcation, and not just from lessons
from Edsger. One phrase from [0] particularly resonates, ``teach each
other,'' which hints at fellowship in the process. Over a sequence of
meetings, advantages of a diverse group of people emerged. Some
sessions were devoted to solving technical problems rather than
reading: then the mix of seasoned researchers, students, visitors,
quick thinkers and slow thinkers contributed to the discussion. Years
later, learning about Kahneman's distinction between fast and slow
thinking made me think back to those meetings.

There was a time, grappling with my own research questions and being
somewhat overcome, I stopped attending. One day, Edsger spotted me in
the hallway and in the kindest possible words, told me that my absence
was missed. Thereafter I resumed regular attendance. Eventually,
techniques I learned from those meetings had cumulative impact on how
I work.  Certainly my dissertation was strongly under the influence of
the Dijkstra style of derivation and proof, though the dissertation
topic was algorithmic and on self-stabilization, another of Edsger's
contributions.

Looking back, I think of three themes from those days.  First is the
utility of a meta-thesis, by which I mean a rough proposal, used
multiple times, in different investigations. My feeling is that for
Dijkstra, this may have been something like a Whorfian hypothesis:
that language, tools, and abstractions shape thought, even in
computing science. When the goal is program correctness, the
hypothesis guides research; we see how a methodology, a discipline for
program construction, and even notation have import.  A second
enduring idea is the avoidance of complexity when it is not
necessary. Even now, when refereeing submissions or examining student
work, I wince a bit when seeing muddled concerns, subscripts that have
subscripts, and so on. It is still true that the hard work of
simplification is underappreciated. A third motif (borrowing a term
coined by Michael Polanyi) is the formalization of tacit
knowledge. This seemed also to be an aspiration of the club, to make
explicit methods of derivation, and further document tools or methods
of formal reasoning.

One final recollection comes from my post-graduate time spent
in Utrecht, where I continued research, but also had the ambition
of learning the local language.  Here Edsger helped me once more:
I corresponded with him in Dutch, and still have hand-written letters.
Eventually I was able to attend a couple of Tuesday Afternoon Club
meetings in Eindhoven (en ook in twee talen$^1$).
\smallskip

\noindent
[0] \href{https://www.cs.utexas.edu/users/EWD/ewd06xx/EWD683.PDF}{EWD683}
To a new member of The Tuesday Afternoon Club.
\medskip

\noindent
$^1$translation: and also in two languages
\hfill 
\hypertarget{linkB}{\hyperlink{linkA}{$\hookleftarrow$}}

 \newpage

\phantomsection
\addcontentsline{toc}{section}{Alain J. Martin}


\centerline{Alain J. Martin, California Institute of Technology}
\centerline{22 February 2021} \bigskip\noindent I arrived in Eindhoven
on a very cold day in January 1971. I had obtained a research
fellowship from the French government and I was on my way to join
Dijkstra’s research group. I had read his article \emph{Cooperating
  Sequential Processes} and was impressed both by the (at the time)
new subject matter ---concurrency--- and by the way in which it was
treated.

His research group at the Technical University of Eindhoven was very
small, just Coen Bron and Wim Feijen. Martin Rem would join a few
months later. It would never be large. The atmosphere was both serious
and cosy, studious and informal. I was immediately invited to his
home. We soon became friends and we would always remain close.

As we all know, Edsger’s contributions were very varied as he had a
talent for identifying important problems and of isolating them in
their simplest and most general form. But a topic that might have been
the single most important red thread throughout his career, was the
mathematical development of programs. That aspect of his work had a
strong influence on my own research. Through his approach, I saw
program notation and the accompanying proof logic become a
mathematical method for developing algorithms rather than simply a
language for describing an already existing solution.

That was the starting point for my work on VLSI synthesis when I
joined Caltech.
I was inspired by Edsger's approach to program
correctness and systematically derived the final design proceeding
through multiple levels of abstraction.  Looking back I am still
amazed that something as complex as a microprocessor can be derived
from a two-page long program by formal transformations.  The
abstractions are so well embedded in the final structure of the
product that no other way of deriving the design would seem
possible. Edsger understood my approach but warned me: ``you will
never convince them $\dots$''

It has been said that geniuses are never one-dimensional in their
talents. It was true of him. First, he had a gift for music. He played
the piano well, and he had a very fine-tuned ear. But mostly he was
extremely gifted for languages. His written Dutch was beautiful, and
we have all appreciated his English in his numerous EWDs and personal
letters. He also spoke German and more French than he was ready to
admit. (He was ostentatiously gallophobic.) If the Dutch I still speak
today has any quality, I certainly owe it to him.

He was always writing, if it wasn't a new EWD, then it was a letter to
a friend, or an entry in his journal. His relation to words was very
special. He claimed, facetiously, that he was in charge of the ‘’Word
Wide Fund’’ whose mission was to salvage words from extinction
$\dots$ In his technical writing, he used language like a precision
tool. For him, precision did not necessarily imply ease of reading,
and he stated that the reader also had to make an effort to
understand. Which once in a while caused friction with some of his
co-authors.

About his attachment to the true meaning of words, he once told me the
following anecdote. He was taking a driving lesson when the instructor
said ``rustig, rustig''. Edsger did not react. Again : ``rustig,
rustig''. At this point, anybody else would have understood that the
instructor meant ``slow down!’’ But in Dutch, ‘’rustig’’ just means
``calm’’ and Edsger was already perfectly calm $\dots$

During Edsger’s lifetime, his country, the Netherlands, went through
truly enormous social and economic transformations. The world in which
the young Dijkstra grew up and the one in which he lived in Eindhoven
were a world-war apart and very different. My guess is that, at times,
he might have felt a little alien in the modern world. He sometimes
gave me the impression that he looked at his fellow human beings as an
outsider. Often, on an evening in front of the fireplace with a whisky
in hand, he would practice his wit with a perceptive and funny
description of human foibles he might have observed recently. He was
sharp but not mean.

Edsger and his wife Ria were the perfect example of the saying that
behind every great man there is a great woman. In my view, he would
not have functioned without her. She was both kind and strong, and
provided the warmth and practical sense he was missing. Faced with a
difficult decision, it was Ria he consulted with, and he always
followed her advice.

He had an aristocratic demeanor. He claimed he believed, quoting
E.M. Foster, in an aristocracy of the mind. That was the quality he
was looking for in his friends, to whom he was very faithful.

Those are the few memories of Edsger W. Dijkstra I chose to share. I
have included some more personal ones, as enough other contributions
will cover his scientific achievements, and I had the privilege of
knowing him well.
\hfill 
\hypertarget{linkB}{\hyperlink{linkA}{$\hookleftarrow$}}



 \newpage

\phantomsection
\addcontentsline{toc}{section}{J Strother Moore}

\centerline{J Strother Moore, Computer Science Department, 
University of Texas at Austin}
\centerline{22 February 2021}
\bigskip\noindent
Dijkstra.  The name is synonymous with precision and analytical thinking
about computing.  His contributions to our science are numerous and lasting.

\noindent
Even his handwriting is famous!

\noindent
But what was the man like?

I first met him in Newcastle, England, probably in 1973, as I was finishing
my PhD at the University of Edinburgh.  I gave a talk on the Edinburgh Pure
Lisp Theorem Prover then being developed by Bob Boyer and me.  Our prover
attempted to find inductive proofs about recursive functions over binary
trees constructed by a pairing function and arithmetic over the naturals
constructed from 0 by successor.  Edsger was in the audience.  And when I
said ``The prover does not support real numbers,'' he raised his hand.  I did
not know who he was but I was taken aback by his question.  He asked ``What
do you mean by real numbers?''  Fortunately, I was using the word precisely
in its traditional mathematical meaning and I described to him a typical
construction of the reals.  After the talk, he invited me to his room where
we drank single malt whisky and he questioned me more deeply about the prover.

He joined the faculty University of Texas Computer Science Department (UTCS)
in 1984.  Boyer and I had joined the department a couple of years prior to
that.  Edsger was familiar with Austin long before he joined UTCS though.  He
frequently visited the Burroughs Research Center in Austin and consulted with
the functional programming language group led by Ham Richards.

When we learned that Dijkstra was considering leaving Eindhoven, the faculty
at UT began to debate whether we should offer him the Schlumberger Chair.  It
is our most prestigious chair.  For most of us the answer was obvious.  Of
course we wanted Dijkstra!  But the decision to offer the chair to Edsger was
not unanimous.

This disturbed and embarrassed me.  I telephoned Edsger in Eindhoven and told
him that the decision was positive but not unanimous.  I did not want him and
Ria to move half-way around the world and then discover that some people
didn't want him there.  But he told me that he was always surrounded by
controversy and that he was used to it.

When he moved to Texas his office was in the UT Tower, where Boyer and I had
our offices, and he almost immediately started the Austin Tuesday Afternoon
Club, which he held in his office every Tuesday for almost twenty years.  He
also frequently dropped in our offices to show us some new proof he had
discovered.  He was always focused on elegance, clarity, and simplicity and
was delighted when he found it.

Our common interests in programming and proof led to many discussions.
But our perspectives were different.  For Edsger, proofs were a way to
explain and understand a computation.  But for Boyer, Matt Kaufmann
(who joined us from Burroughs) and me, mechanically checked proofs
were ``just'' a way to recognize logical truth.  For example, what
role does case splitting play in proof?  Edsger disliked case
splitting while we allowed our prover to use it fairly freely: humans
often get lost in big case splits while machines manage them well,
within limits.

He taught an undergraduate course on the logic of programming.  I
frequently taught in the same classroom he did, right after his class.
He would always write something on the board for his students.  My
favorite was a quote often attributed to Bertrand Russell, ``Most
people would sooner die than think. Many of them actually do so.''
                   
His honesty and insights could be maddening.  A faculty member once said to
me ``Until I got to know Edsger I never understood why they poisoned
Socrates.''  And I have met many people from other universities who have said
``It must be horrible having Dijkstra on the faculty.''  But I served as
department chair from 2001 to 2009 and I wished I had 10 more like him.

It was in faculty debates ---usually over hiring and promotion--- that Edsger
stood out most.  He always argued for the best and the brightest.  He did not
care so much what field a person worked in, as long as the person was working
clearly on deep questions.  He was surprisingly compassionate about junior
colleagues.  Many times I heard him say that tenure would help a young
researcher break loose.

He never threw his weight around.  He never came into my office after
a meeting to suggest that I should do things his way rather than the
way the faculty had voted.  He would state his opinions in the open,
debate the issues, cast his vote, and then go to lunch with whomever
wanted to eat with him.  He did not conspire with other faculty to get
his way and, to my knowledge, he never went to the Dean or other
higher university officials ---where his international stature would
have surely gained him an audience and sympathetic ears--- to try to
bend the department to his will.

Edsger loved his adopted state and country.  He came to work in the summer
wearing a big Texas cowboy hat ---they are made to keep you cool in the Texas
sun.  Often he would have on a cowboy's string tie.  So from the waist up, he
looked more Texan than I did.  But he almost always wore shorts and sandals,
which ruined the cowboy image completely.

He and Ria loved to travel.  They explored the American continent.  I once
bumped into them in the high desert of West Texas, exploring Big Bend
National Park.  They had an RV ---a ``recreational vehicle''--- a big house on
wheels that they drove all over the country.  They called it the Touring
Machine.

Edsger was like a man carrying a light in the darkness.  Almost every time he
said something, the issues were illuminated.  At faculty meetings when Edsger
was not present it was common for someone to say ``If Edsger were here he'd
say such-and-such.''  That was just another way to say ``The right thing to
do, politics and personalities aside, is such-and-such.''

I learned from him every time I interacted with him.  And it was no different
at the end.

He was very matter-of-fact about his cancer and impending death.  Soon he was
in hospital and the doctors predicted he did not have long to live.  I
visited him several times and we talked about proofs, about the
department, about how to be a professor, about politics, about science.

I visited him one last time just before he got on the plane to return to the
Netherlands.  It was obvious to both of us that it would be our last meeting.
When it was time for me to go, I said good bye and I told him that I would
miss him.  He looked me ---with a twinkle in his eyes--- and said ``Well, I
won't miss you.''  We were both just being honest with each other.
\hfill 
\hypertarget{linkB}{\hyperlink{linkA}{$\hookleftarrow$}}

 \newpage

\phantomsection
\addcontentsline{toc}{section}{Vladimir Lifschitz}

\centerline{Vladimir Lifschitz, Computer Science Department, 
University of Texas at Austin}
\centerline{23 February 2021}
\bigskip\noindent
Friendship with Edsger and Ria was a wonderful gift that Austin gave
us when my wife and I moved here in 1991.  Losing them many years
later was a great personal loss.

Edsger was interested in ``streamlining'' mathematical arguments, and
his views on the organization of proofs had a profound effect on my
professional work. As an undergraduate, I had learned that proof can
be best understood as natural deduction—introducing and discharging
assumptions.  Conversations with Edsger convinced me that, in many
cases, it is better to present a proof as a chain of equivalent
transformations.  As an example, Edsger took the list of theorems that
students in my logic class had been given as exercises on the use of
Peano axioms, and showed me how to prove them in the Dijkstra/Scholten
``calculational style.''  The proofs were concise and elegant, like
every other product of his thought.

This was an eye-opener. Examples of calculational proofs in Edsger’s
writings were so impressive that I even asked myself whether every
possible use of natural deduction in classical logic can be replaced,
in principle, by calculational reasoning.  The answer turned out to be
yes (published in the \emph{Annals of Pure and Applied Logic} in 2002).

Using simple, economical notation is an important rule of mathematical
writing that I learned from Edsger.  No unnecessary subscripts!  One
day he showed me a place in a draft that I had asked him to review,
where formulas included (I am ashamed to admit) two levels of
subscripts, and said: ``I showed this page to my students as an example
of how NOT to write mathematics.  I didn’t tell them, of course, who
the author is.''

I cannot say though that my current views on mathematical reasoning
are completely in line with Edsger’s.  He did not approve of using
pictures, and I learned that from our very first conversation about
mathematics. Prior to applying for a faculty position at the
University of Texas, I came to Austin on an exploratory visit, and
Krzysztof Apt invited Edsger and me for dinner.  Edsger offered me a
tricky puzzle (which is discussed, as I learned much later, in
\href{https://www.cs.utexas.edu/users/EWD/ewd10xx/EWD1067.PDF}{EWD1067}).
In his eyes, that was probably part of the forthcoming job interview.
My solution used a graph that I sketched on a paper napkin.  Edsger
said that he did not like my geometric approach, but admitted that the
answer was correct.

In spite of committing such a grave sin, I was offered the position,
which could not have happened without Edsger's endorsement.

Now that Edsger is not with us anymore, I often remember him when I am
reading or writing mathematics.  I tell myself, ``Edsger would have
expressed this in a different way $\dots$''
\hfill 
\hypertarget{linkB}{\hyperlink{linkA}{$\hookleftarrow$}}

 \newpage

\phantomsection
\addcontentsline{toc}{section}{Wim H. Hesselink}



\centerline{Wim H. Hesselink, University of Groningen}
\smallskip
\centerline{whh572, 24 February 2021}
\bigskip
\noindent
The first time I met Dijkstra was on Tuesday 16 July 1985.  This
meeting came about as follows.  I had got my doctorate in pure
mathematics from the University of Utrecht in 1975.  The next year I
moved to Groningen.  In 1983, I had lost sight of the research front
in my branch of mathematics. Around the same time Jan van de
Snepscheut, a former student of Dijkstra, was appointed professor in
computing science in Groningen.  He turned out to be an inspiring
leader of an emerging research group.  This gave me confidence to move
to computer science.  To finalize this move, the Institute granted me
a sabbatical year with Dijkstra at the University of Texas at Austin,
where Dijkstra had been appointed in 1984.

The meeting with Dijkstra took place in Nuenen.  He had announced to
have a beard and sandals, and a VW Sirocco.  I had not expected him to
wear shorts.  He spoke softly.  I often was thinking to raise a new
topic, when he proceeded with the previous one.  Jan had prepared me
for this by admitting he found it difficult to talk to Dijkstra over
the phone.  After explaining my switch to computer science, I started
with the remark that my main difficulty in the new field was to decide
which things were important.  Dijkstra seemed to approve this, and
came with an anecdote about certain logicians who confronted with
linear search, asked about the computability of the search criterion.
He explained that the central problem of computing science was to
restrain the complexity of our artefacts, and not to make a mess of
it.  I tried in vain to seduce him to divide computing science in
subdisciplines, as I knew mathematics could be divided.

From September 1986 until May 1987, I worked in the second room of
Dijkstra's office in Austin.  My weekly highlight was the Austin
Tuesday Afternoon Club (ATAC), where Edsger always could inspire a
small group of colleagues to investigate and discuss some problem.
The first problem was a preprint by Greg Nelson about Dijkstra's
calculus. We concluded this investigation in two or three sessions,
and found some minor mistakes.  Edsger strode to the phone and
informed the author about our findings.  This way of compromising the
anonymous reviewing system was new for me.  My main learning
experience was not in the contents of Nelson's paper but in the way
Dijkstra and the others appreciated it.  At the end of my year in
Austin, there was an ATAC session in which Pnueli was invited to
explain his temporal logic.  He had difficulty to defend the relevance
of his logic against the sceptical attacks of Dijkstra and Hoare.
Indeed, Hoare was that year also in Austin, and often attended the
ATAC.  This was helpful, because Dijkstra alone could have been
overwhelming.

Next to the ATAC, I attended an undergraduate honours course Edsger
gave on mathematical methodology.  Quite strange, since I had a
doctorate in math, and knew it much better than Edsger.  Yet I
appreciated this course and had the impression I learned things.  Once
in the course, I used hands while explaining something.  Edsger then
told me to try to do it without hands.  Some weeks later, I saw Edsger
using hands to explain the lexical order, but I kept this to myself.
During this year, I worked on several ideas suggested by Dijkstra or
Hoare.  Invariably, Dijkstra's suggestions turned out mathematical,
while Hoare's ideas involved languages.

The sabbatical year with Dijkstra has broadened my view of computer
science immensely, because of the contact with Dijkstra, Hoare, and
other colleagues, and because of the string of conferences of the Year
of Programming that was organized in Austin.  The direct influence of
Dijkstra was limited.  Before coming to Austin, I had read and used
\href{https://www.cs.utexas.edu/users/EWD/ewd08xx/EWD883.PDF}{EWD883}
(by Dijkstra and Scholten), but when I told Edsger this, he said he
did not like it, presumably because it was too operational.  Another
influence was that afterwards I have organized in Groningen, for 20
years, a reading group, with some staff and master students, as a weak
simulation of the ATAC.

In subsequent years, Edsger and I mainly corresponded by letters.  In
1988, I attended the Summerschool at Marktoberdorf, with Dijkstra as
one of the directors.  Between the sessions he was often surrounded by
a group of disciples.  In 1989, he attended the first MPC conference
in Twente, which was organized by our Institute.  In March 1991, I
suggested him by letter to collaborate on something like concurrency.
He declined gracefully.  About concurrency, he wrote: ``I don't like
it, because it is such an operational concept \dots''.  In the winter
of 1994, we were all deeply shocked by the tragic death in California
of Edsger's beloved disciple Jan van de Snepscheut.  Edsger and I
extensively communicated about Jan's history in Groningen.

In 1990, Edsger received for his 60th birthday a Festschrift titled
``Beauty is our business''.  This title was a quote from
\href{https://www.cs.utexas.edu/users/EWD/ewd06xx/EWD697.PDF}{EWD697}.
It expressed Edsger's view of computing science.  I have tried to live
up to it, but I often failed, either because of lack of creativity, or
because I did not have the time.  In my contribution to the
Festschrift, however, I thought I succeeded.

Edsger encouraged his followers to imitate his behaviour. I was too
old to comply, but I had a beard when we first met, and my handwriting
was acceptable though quite different from his own.  Edsger respected
my independent intelligence, and we shared the love for a convincing
mathematical argument.  Anyhow, following Dijkstra with his EWD
series, I started to enumerate my manuscripts as whhxxx. I still do
this, though not as rigid as Edsger would have wanted.  This is one of
the few lasting influences. In spring 1987, around the time of
\href{https://www.cs.utexas.edu/users/EWD/ewd10xx/EWD1005.PDF}{EWD1005},
Luca Cardelli had distributed a pamphlet marked EWD1024, in a font
that emulated Edsger's handwriting.  Edsger and Ria were not amused.
Eleven years after his death in 2002, I met a brother of Edsger.  When
I told him I had been with Edsger for a year, he looked at my feet and
asked: ``Why don't you wear sandals?''
\hfill 
\hypertarget{linkB}{\hyperlink{linkA}{$\hookleftarrow$}}

 \newpage

\phantomsection
\addcontentsline{toc}{section}{Hamilton Richards}


\centerline{Hamilton Richards}
\centerline{28 February 2021}
\bigskip\noindent
In 1976, a freshly minted PhD looking for a job, I received an offer
from Burroughs Corporation. Its B5000 series had earned the approval
of the acclaimed Turing Awardee Edsger W. Dijkstra, who was the
company’s Research Fellow. This convinced me to join a Burroughs
laboratory in San Diego which was investigating data-flow computing.

Edsger’s fellowship entailed consulting visits to Burroughs plants and
labs. During his first visit to the San Diego lab after my arrival,
Edsger noticed a binder in my bookshelf, labeled “E.W. Dijkstra,”
containing photocopies I had received as a graduate student. He
remarked, “This is how authors are cheated out of their royalties.” It
was a rough start for a friendship whose significance in my life was
exceeded only by my marriage to Joanne.

In 1978 the data-flow project was canceled, and I moved to Austin as a
founding member of the Burroughs Austin Research Center (BARC)
investigating functional programming (Robert Barton, the B5000’s chief
architect, had been inspired by the Church-Rosser theorem’s
implications for concurrent processing).

During Edsger’s first Austin visit he complained about his hotel
room’s lousy writing table, so I invited him to do his writing at our
house. In subsequent visits, he accepted our invitation to stay at
`Hotel Richards'.  Conversations at dinner and into the evening were
engrossing, ranging widely over politics, culture, academia,
technology, and the English language. He became a fan of
\emph{The New Yorker}—especially the cartoons. We fondly remember Edsger stretched
out on the living room carpet, reading contentedly while our
5-year-old son covered him with sofa cushions.

At BARC we had lively discussions with Edsger about proving
correctness of purely functional programs (e.g.,
\href{https://www.cs.utexas.edu/users/EWD/ewd08xx/EWD825.PDF}{EWD825},
\href{https://www.cs.utexas.edu/users/EWD/ewd08xx/EWD827.PDF}{EWD827}). At
first Edsger contended that although functional programs are typically
much shorter than their imperative counterparts, their correctness
proofs would be longer. Further investigation refuted this
supposition, and in later years Edsger’s appreciation of functional
programming was revealed in a memo [0] defending the use of the
functional language Haskell in my introductory programming course.

One of Edsger’s visits coincided with a visit by David Turner, who was
consulting for BARC. Edsger was in the audience for one of David’s
lectures, and in the style for which he was known, he began to
interrupt with questions and objections. Given the two towering
personalities, escalation was inevitable, and finally David aborted
his lecture. At dinner that night, Edsger conceded that he’d gone a
bit over the line, and in subsequent days at the lab he and David got
along well.

BARC had a no-smoking policy, but recognizing Edsger’s smoking habit
we set up an office for him with special ventilation. His response was
a blistering memo attacking society’s growing disapproval of
smoking. A few years later he quit smoking, so inconspicuously that
three days passed before his wife, Ria, noticed.

On every visit to Austin, Edsger was invited to give lectures at the
University of Texas. As Burroughs support for his work declined, the
visits began to include interviews, and in 1984 he was appointed to
the Schlumberger Centennial Chair in Computing Science.

That meant moving to Austin and finding a house. Edsger’s stays at
Hotel Richards having familiarized him with the neighborhood, he and
Ria chose a house just up the hill from ours. It was a new house,
requiring modifications including one to make space for the
Bösendorfer grand piano. Staying with us for six weeks, they became de
facto members of our family.

Edsger’s love for Texas—the wide blue skies, the friendly Texans, and
the landscapes’ variety—is captured in the T-shirt he wears in the
enclosed photo. He and Ria toured widely in their VW camper, dubbed
“Harvey the RV” or “the Touring Machine.” Staying in state parks, Ria
would cook while Edsger would write. Edsger loved Texas barbecue, and
we dined often with them at the nearby County Line restaurant.

Their regular after-dinner walks in the neighborhood frequently passed
by our house, and they would stop in several times a week for
conversation, iced coffee, and—for Edsger—a glass of scotch whisky.

In 1986 BARC closed, and Edsger recommended me to coordinate UT’s Year
of Programming, a US Navy-sponsored series of workshops. That was a
half-time position, so I was assigned to teach for the other
half. Thus began my 18-year career as a Senior Lecturer. The subjects
I covered in my courses included Edsger’s calculational reasoning and
weakest-precondition methodologies, but his most important influence
on me is summed up in his epigram:

\begin{quote}
$\dots$ if 10 years from now, when you are doing something quick and
dirty, you suddenly visualize that I am looking over your shoulders
and say to yourself, Dijkstra would not have liked this, well that
would be enough immortality for me.
\end{quote}

After years of digestive problems, in early 2002 Edsger was found to
have esophageal cancer. Out-patient treatment failed spectacularly—in
his first day home from the hospital, wearing a chemotherapy pump, he
collapsed so violently that his head made a huge dent in the bathroom
wall. After weeks in hospital, he was pronounced (barely) fit to
travel. Home for a day, he and Ria hosted a crowd of visitors at lunch
with barbecue from the County Line, and he played a little Mozart on
the Bösendorfer. Then they flew home to Nuenen, accompanied by me as
logistical assistant.

Returning to Austin, I undertook the task, with help from my wife and
Jay Misra, of closing up the Dijkstras’ Austin household. What was not
sold was shipped to Nuenen (including the Bösendorfer). The job was
done in time for us to leave for our summer vacation in New
Hampshire. I neglected to take my passport, so that when the word came
that Edsger had died, there was no way for me to get to Nuenen in time
for the memorial service. The regret lingers on.
\smallskip

\noindent
[0] \href{https://www.cs.utexas.edu/users/EWD/OtherDocs/To%20the%20Budget%20Council%20concerning%20Haskell.pdf}{To the members of the Budget Council}, Edsger W. Dijkstra, 12 April 2001.

\smallskip

\centerline{\includegraphics[width=0.79\textwidth]{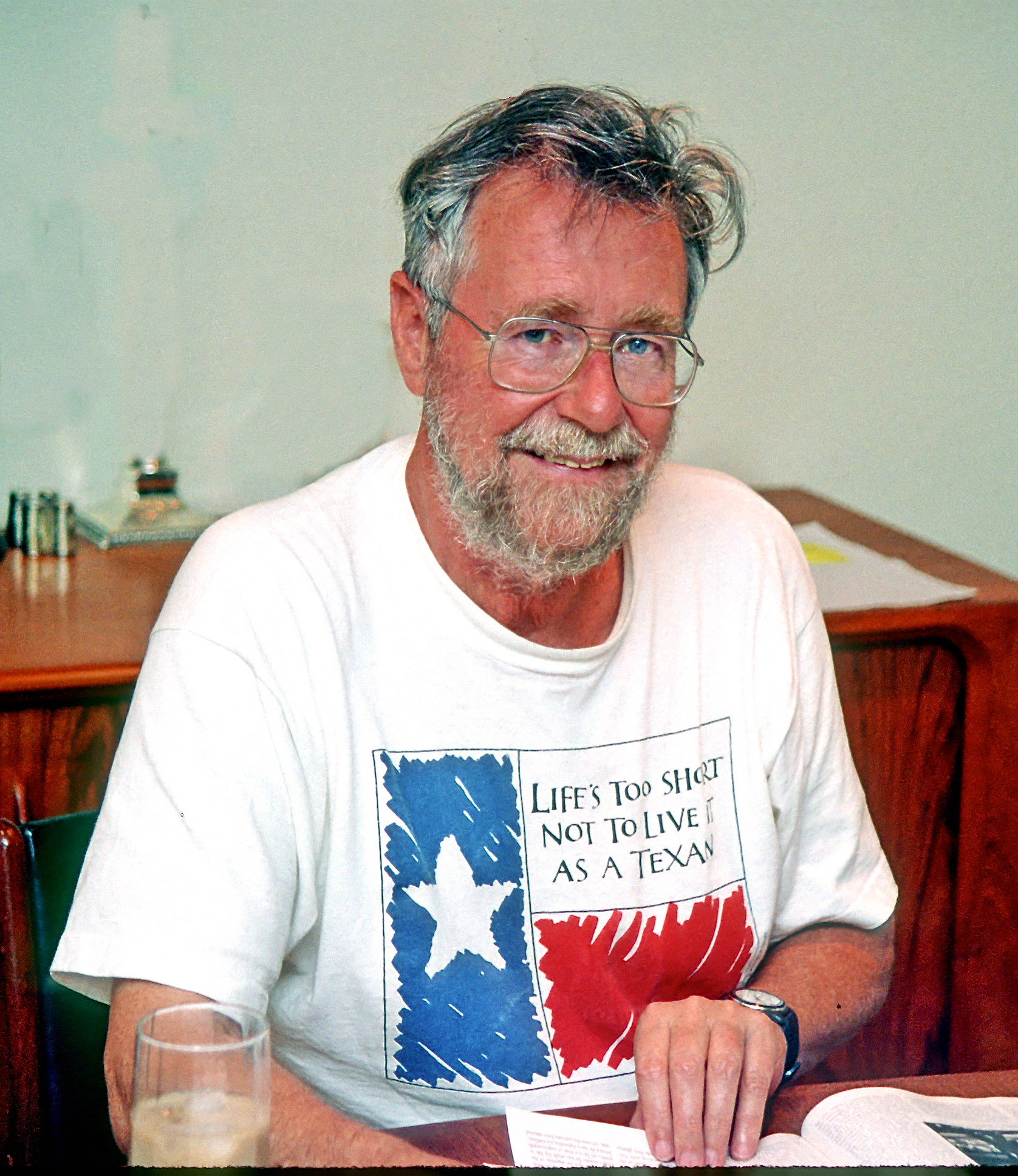}}
\hfill 
\hypertarget{linkB}{\hyperlink{linkA}{$\hookleftarrow$}}


 \newpage

\phantomsection
\addcontentsline{toc}{section}{Ken Calvert}



\centerline{Ken Calvert, Computer Science Department, University of Kentucky, Lexington}
\centerline{28 February 2021}
\bigskip
\noindent
Dear Edsger,

When I arrived in Austin to begin PhD studies in
the summer of 1984, you had just joined the University of Texas (UT),
but the mythology around you had already started to grow.
There was a rumor among students that your office, on the 21st
floor of the Texas Tower, housed a grand piano that had to be hoisted up
the outside of the building because the elevator was not large enough for
it.  The graduate students knew---and this was no myth---that yours
was not an ordinary class, including as it did the prospect of an oral
final examination one-on-one with a Turing Award winner.

I knew of your eponymous algorithm and of semaphores from my prior
studies, but I was not familiar with your other work.  One day that
summer, as I was browsing in the UT bookstore, I happened across your
book \emph{A Discipline of Programming}---probably in a display of
books by ``local authors''.  I bought it, and it changed the way I
think about programming.  I had considered myself a pretty good
programmer, and had taken at least one CS theory course, but somehow
it had never ``clicked'' for me that \emph{real\/} programs were
amenable to formal mathematical reasoning and manipulation.  That was
probably the first of the many things I learned from you.  Though you
were not my advisor, you had a great influence on me during that
formative period of my career---through your classes, your writings,
and the people around you.
Today, scarcely a week goes by when I don't
quote something you said, or refer to something I learned from you or from
the Austin Tuesday Afternoon Club (ATAC).  I'm pretty sure I have not
always done those things justice, but I hope that your
wisdom has, in some small way, been reflected to those in my own
circle of influence over the years.

It took me a year or two to muster the courage to take your 
course, and when I did, I took it ``pass-fail''.  (For those not familiar
with the American educational system, that is a lower-stress
option compared to the normal system, in that it does not distinguish
among levels above the threshold required to pass.)
When you subsequently
invited me to join the ATAC, I was both thrilled and intimidated.  (I
learned then that, by that time anyway, there was no grand piano in
your office.)

In your classes and in the ATAC, besides learning about mathematics
and programming, I learned much about how to think,
speak and write precisely---and also about the importance of doing so.
Among the principles, wisdom, and values I got from those experiences,
Rule~0 (``Don't make a mess of it'') is one that all my
students, not to mention my kids, have heard. Occasionally, when it
seems especially appropriate, we even recite it together as a class.  
Among other pearls I picked up that are still with me, I would mention these:
\begin{itemize}
\item
Notation matters. Some ``standard'' mathematical
notations are not only not helpful, but actually hinder
understanding.  An hour spent refining notation to save a hundred
readers a minute of thinking is well spent.

\item
The purpose of a calculus is to \emph{let the symbols do the work}---to
the greatest extent possible.
I still use the calculational proof style you taught in your class and
used in your book \textit{Predicate Calculus and Program Semantics}.

\item
``If you must choose between beauty and utility, choose beauty,
because the world has enough ugly useful things.''  (The last part is
at least as accurate now as it was then.)

\item
Integrity.  I think all teachers convey values to their students to
some degree, but I had never met someone
with such strongly held convictions, as well as the
intellect and confidence to defend them---and even to call out some
who disagreed in published writings.
(Your \textit{Selected Writings on Computing\/} was another eye-opening
book for me.)
Observing your interactions with the brilliant and famous people who visited
Texas and sat in your class or the ATAC was always instructive.
Sometimes those interactions were cordial, sometimes not. (I recall
you sitting in the back of the room during a talk by one distinguished
visitor and making rude noises).  Today, communicating values is
explicitly part of my teaching philosophy, though I am not
as fearless about it as you were!
\end{itemize}
These experiences raised my consciousness in many ways. That comes with
a price, especially for those with perfectionist tendencies: one must
be able to meet one's own standards, at least most of the time, in
order to be productive. While this has sometimes been a challenge for
me, I would not have it any other way.

You helped me along in my career, enabling me to attend a
NATO Summer School in Marktoberdorf, serving as a reference in my job
search, and staying in touch for quite a few years.  You were kind and
generous to me and my family personally while I was still a student,  stopping
for a beer and a visit when we chanced to meet each other while camping
in one of Texas' state parks (you and Ria in the ``Touring Machine'');
inviting my wife and I to dine with you and Ria in your home as I
neared completion of my studies.  On that occasion, I admired the design of a
Brabantia corkscrew you were using.  You said ``Wait a bit,'' rummaged
in a cabinet for a few moments, pulled out one just like it---brand
new and still in its package---and presented it to me. 
Like so many things great and small that I received from you, more than
thirty years later I am still using that corkscrew.  For all of those
things, I thank you.

The enclosed photo was taken in the Calvert home, during a visit by
EWD to Atlanta sometime in the mid-1990's.  (I believe he had given an
invited talk at Emory University.)

\smallskip

\centerline{\includegraphics[width=1\textwidth]{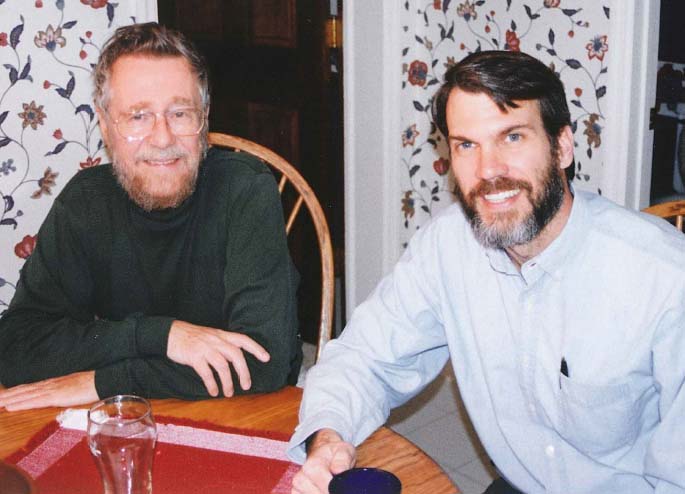}}
\hfill 
\hypertarget{linkB}{\hyperlink{linkA}{$\hookleftarrow$}}

 \newpage

\phantomsection
\addcontentsline{toc}{section}{David Naumann}
\centerline{David A. Naumann, Stevens Institute of Technology, Hoboken NJ}
\centerline{1 March 2021}
\bigskip
\noindent 
Around 1994 when I stayed with Ria and Edsger in Nuenen he took me to
a shop in Amsterdam for repair of his fountain pen.  That evening he
gave me a pen and I said my writing was not worthy.  His response, in
friendly tone: ``It may improve.''  This was neither the first nor the
last time I benefited from his faith in people's ability to change.

One of my first personal encounters with Edsger, around 1987, was an
opportunity to change my mind: He told me he and Ria decided to move
to Texas to show they could learn new tricks.  That made me warm to
him, overcoming a previous dislike that can be blamed on free pizza
when I had just started graduate study at UT.  At that time (1984) I
was taking courses to compensate for my minimal undergraduate training
and I knew next to nothing about the person who gave a dinner speech
at an ACM event.  His ranting about the poor quality work being done
on his new home made me dismiss him out of hand.  My negative
impression was reinforced when I started attending department
seminars: UT had recently banned smoking on campus and Edsger was the
lone scofflaw among many dozens of students and faculty.

A self-supported PhD student was allowed to be rudderless, and at
first I was.  Then in Fall 1986 friends steered me to take a course
with Tony Hoare, who advised me to take Edsger's course, which changed
my life.  Through Edsger and Tony I came to understand programming,
which for me had been an enjoyable and lucrative craft, as a
scientific discipline.  I became enamored of the possibility to
derive, by calculation, correct solutions to programming problems.

Edsger showed the efficacy of manipulating symbols by formal rules
without thinking much about the interpretation of those symbols.  He
also made pronouncements about being a formalist, which sounded
dogmatic and didn't tell the whole story.  During one of the events of
the Year of Programming (1987) I had the pleasure of sitting at lunch
with Edsger and Bob Boyer who was trying to get Edsger to admit the
existence of epsilon-nought, an ordinal number important in
constructive mathematics.  Bob sketched increasingly complicated
``bags'' on a napkin.  Edsger seemed open-minded, not at all dogmatic,
and was friendly to me.  It was my first hint that some of Edsger's
public behavior was theatrical, drawing attention to himself and his
ideas, backed by profound dedication to science and compassion for
students.

The following year, in Edsger's course, I complained that his
presentation of predicate calculus allowed nonsense like the predicate
of Russell's paradox.  He said it was an issue he chose not to
address.  He advocated a strictly formalist position yet seemed to
disdain type theories that formalize constructivist mathematics in
avoidance of contradiction.  Later I came to appreciate what I
understand to be an engineer's use of mathematics: We can rely on our
interpretations to keep us from nonsense, while getting the most
benefit from symbols by neither cluttering the rules nor thinking
about interpretation.  Edsger's performances in class were stimulating
and it was abundantly clear he respected and cared about students.  I
was shy and insecure, though, and when it came time for the oral exam
I did not do well; yet Edsger was patient and warm during the entire
conversation. I left feeling encouraged.

By the time Edsger quit smoking I was participating in the Austin
Tuesday Afternoon Club (ATAC).  The force of his will to change was on
display when he quit cold turkey.  In the ATAC I often saw that
Edsger's strongly held views were open to revision and refinement.  He
criticized conventional practices in mathematics and logic, but one
afternoon after coming to appreciate a particular fine point he said
to Allen Emerson, ``you logicians are no fools.''

Tony had posed the problem for my dissertation: category-theoretic
laws of predicate transformers. Edsger, Ralph Back, and Carroll Morgan
had made clear the value of predicate transformers as a uniform basis
for calculating with specifications and programs.  Generalizing their
work to higher order programs seemed fraught with the danger of
logical inconsistency so I spelled out set-theoretic interpretations
of everything, in pedantic detail and in contrast to the axiomatic
style Edsger was using to develop predicate transformer theory in
collaboration with Carel Scholten.  Complicated category-theoretic
definitions were not to Edsger's taste; he once proposed that I write
a paper on why theoretical computer scientists do not need to know
category theory.  Nonetheless, when it came time to review my draft
dissertation (over 200 pages), Edsger read almost every word and
formula, as evidenced by extensive margin notes in pencil.  He invited
me to his home and we spent many hours going through the text and
discussing it in detail.  There were only a couple of sharp comments;
in connection to some dense and wordy summary paragraphs he wrote,
``This is a style of doing mathematics that I abhor.''  But I had done
a lot of calculations and he wanted to understand everything.  Rather
than dwell on whether my general approach was a good idea, we
discussed the rationale for various design decisions, technical
details, and notations.  He guided me to many corrections and
improvements.

Through Edsger's example I came to appreciate the importance of the
university and the ways academic scientists can contribute to society.
I took up that path myself and as we became friends Edsger continued
to accept me as I was, while inspiring me to care for others and
helping me improve.  He introduced me to the term doctor father and he
was one to me.
\hfill 
\hypertarget{linkB}{\hyperlink{linkA}{$\hookleftarrow$}}


 \newpage

\phantomsection
\addcontentsline{toc}{section}{David Turner}


\centerline{David Turner, University of Kent, UK}
\centerline{1 March 2021}
\bigskip
\noindent
My personal encounters with Edsger Dijkstra date from 1980 and began
by letter. I received a note from Edsger dated 7 May 1980 enclosing a
copy of
\href{https://www.cs.utexas.edu/users/EWD/ewd07xx/EWD735.PDF}{EWD 735},
``A mild variant of combinatory logic'', which he sent
to me at the suggestion of Ham Richards, relating to my January 1979
article in Software Practice and Experience. Edsger had been
sufficiently interested in the topic to set about inventing his own
version of combinatory logic as a way of better understanding it. In
the note he also enquired if I would be at the Burroughs facility in
Austin, Texas in August when he expected to be there.

I was to meet Edsger much sooner. I had been invited to give seminars
at six Dutch computer science departments on a visit organised for me
by Doaitse Swierstra at Groningen, and on Tuesday 20 May, I found
myself addressing Edsger’s Tuesday afternoon club at Eindhoven. After
the discussions had finished, Edsger insisted on my staying over with
himself and Ria at their house in Nuenen. They were gracious hosts
there, as I found them to be later in Austin, and forty years on I
recall the evening with pleasure. They had two very large dogs,
creatures of which I am normally wary, but the Afghan wolf-hounds
fortunately proved docile. Edsger enquired about my family background
and appeared satisfied to discover that my father was a
businessman. Edsger was interested in people. His trip reports
frequently contain acute sociological observations, and while famously
intolerant of what he judged bad science or sloppy reasoning, Edsger
was in my observation never an intellectual or social snob.

In the morning I saw Edsger sprinkle on his breakfast brown powder
from a tin whose label said, according to my Dutch pocket dictionary,
“ground mice” which I had to ask about. This turned out to be powdered
aniseed. Another idiosyncracy which I first noticed on that occasion
was that Edsger wore two watches, one on each wrist. One, I think the
left, was set to the local time of whatever time zone he was in, the
other was always on Dutch time. Edsger referred to this, humorously,
as “God’s time”, although from the indications available to me and
reported by others I am fairly sure Dijkstra was a non-believer. He
was, however, very definitely Dutch.

I met Edsger again that year, in August, at Burroughs Austin Research
Center, as his letter had anticipated. He took an immediate interest
in SASL, a simple lazy functional language which Burroughs had adopted
for the project at BARC, and set a problem which we solved together,
to generate the decimal digits of ``e'' as an infinite list. Edsger
contributed a crucial lemma, without which my program could not have
worked. He wasn’t comfortable with ``infinite list'' by the way,
preferring “potentially infinite” or something similar. I wondered if
he had been influenced by the intuitionist school of mathematics,
whose members reject the idea of already completed infinities. Years
later, I had the opportunity to ask Dijkstra if he had known Brouwer 
---apparently they did overlap, both being members of the Dutch Academy
of Sciences.  Edsger said that Brouwer was the most argumentative
person he had ever known (these may not have been the exact words but
that was the sentiment) which coming from Edsger was a strong
statement.

I had the privilege of interacting with Dijkstra quite often in the
period from 1980 to 1984 when he ceased to be Burroughs Research
Fellow and took up a Chair at the University of Texas. I was a
consultant to the functional programming and combinator reduction
machine project at Burroughs Austin Research Center from January 1980
until it was, sadly, closed in 1986 following the merger of Burroughs
with Sperry.  Edsger took a definite interest in this project and
sometimes became quite involved during his visits. I recall that Mark
Scheevel devised a better method of extracting combinators from SASL
``where'' expressions, which made it practical to use a copying version
of the ``Y'' combinator in place of a cyclic one, and thus a reference
count garbage collector. Edsger became really excited at this
development. He was certainly interested in the practical problems of
computer engineering, as is apparent from chapters in his career.

I was able to interact with Dijkstra from time to time in the years
that followed but regrettably much less often than in the early
‘80s. I include a photo of Edsger and me in conversation at the
Barmitzvah of Ben Richards (Ham and Joanne’s son) in 1990. I can
recall another occasion when I dined with Edsger and Ria, this time at
their house in Austin. I was a visiting professor at UT for the Spring
semester of 1992, to which I had come without my family. My wife had
given me a rather nice, remarkably small, portable CD player which I
thought would interest Edsger, which it did. He insisted that I borrow
whatever I liked from his extensive classical music collection.

Edsger was a truly unusual person. He had many idiosyncracies which
will doubtless be mentioned by others: his insistence on numbering
from zero; his dislike of canned music ---which he would sometimes take
direct action to eliminate at source with varying results as I
witnessed on several occasions; making his own ink. But for me his
most striking property was absolute intellectual honesty. I have an
invisible Edsger inside my head which looks over my shoulder when I am
writing and quietly goes ``Tut tut'' if I write something that is
muddled or not accurate. I don’t always listen to that voice but know
I should. Some thought Edsger arrogant. He was dismissive of work he
thought unworthy of attention, a category that for him was quite
large. But to achieve as much as he did probably requires the ability
to cut out noise. For someone of such high intellectual gifts Dijkstra
was remarkably modest.
\smallskip

\centerline{\includegraphics[width=1\textwidth]{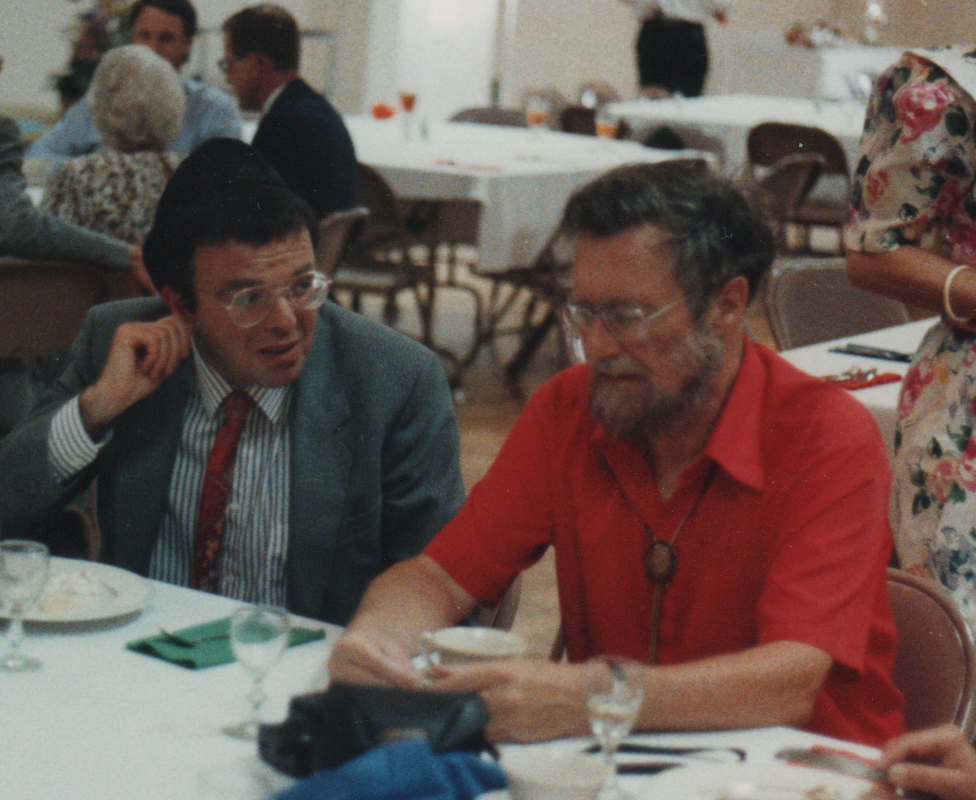}}
\hfill 
\hypertarget{linkB}{\hyperlink{linkA}{$\hookleftarrow$}}

 \newpage

\phantomsection
\addcontentsline{toc}{section}{J.R.~Rao}




\centerline{J.R. Rao, IBM Thomas J. Watson Research Center, Yorktown
  Heights, NY} \centerline{1 March 2021}
\bigskip
\noindent
My first awareness of Prof. Edsger W.~Dijkstra came when I was an
undergraduate freshman at the Indian Institute of Technology,
Kanpur. His reputation as a deep and influential thinker had preceded
him across the globe. As an aspiring computer scientist, I purchased a
copy of his classic book entitled, \textit{A Discipline of
  Programming}. As I had been forewarned, I found his writings to be
abstruse, requiring a maturity that I lacked. Even so, in a
premonitory way, I left the book on my shelf till I came back to it
many years later.

In the Fall of 1986, I joined the doctoral program at the University
of Texas at Austin. I was excited as the faculty roster comprised
several pioneering, world-famous researchers and the department had
just launched the Year Of Programming. I first ran into Prof.~Dijkstra
as he was waiting outside Dr. K. Mani Chandy and J. Misra's offices,
in his signature Birkenstock sandals, with his long leather purse,
smoking a cigarette. I felt intimidated to see him in person and he
barely acknowledged my greeting. Sensing his preoccupation and
suspecting aloofness, I left.

As part of graduate coursework, I took three consecutive courses
taught by Prof.~Dijkstra entitled \textit{Capita Selecta} -
\textit{Selected Problems in Computing}. In the very first class, he
asked us to write the English alphabet (in both upper and lower cases)
and the arithmetic digits and having done so, to reflect if we could
distinguish between the lower case 'p' and upper case 'P', between the
upper and lower case 'O' and the digit '0', between the lower case 'l'
and the digit '1' in our hand-writing. Thus began a journey in
learning how there was no detail that was too minute for our
consideration and how no effort would be spared in order to not
confuse our readers and ourselves.

Prof.~Dijkstra's classes were a joy to attend. For each class, he
would chose a programming problem and challenge us to share our
solutions on the blackboard. He was a kind and gentle teacher who
would guide us by suggesting changes to the presented solution.~Did we
choose our notation correctly for framing the problem and was that
particular subscript really necessary? He would repeatedly stress the
importance of choosing words carefully: ``If you have to use your
hands, then there is something wrong with your words'', he would say,
adding ``imagine there is a blind man in your audience, how would you
speak?''

Over time, I came to learn and appreciate that Prof.~Dijkstra's class
was operating at two different planes. There was the lesson and then,
the lesson within the lesson. At one level, the goal was to solve the
presented problem. At the second and more richer meta-level was the
approach for arriving at the solution. Many have commented on his
methodical and economic use of the blackboard in his classes. Despite
the complexity of problems that were tackled, he was somehow able to
fit the solutions in one blackboard. I began to appreciate how careful
choice of powerful notation and tools could help us master complexity
and communicate our arguments crisply. Prof.~Dijkstra taught us that
in computing science, complexity comes for free; one has to work hard
for simplicity. Others commented on how he would not react to
questions immediately, but wait for a minute or two before
responding. I came to appreciate his thoughtfulness as he would
examine both the source of the question and confusion before giving a
measured reply. In this way, the classroom became a forum for a
generation of computing scientists to examine their instinctive
approaches to problem solving and revisit their instinctive methods of
reasoning and communication. Our final course grade was based on a
one-on-one, face-to-face, final examination with him, conducted using
a pen and a few sheets of white paper where he could observe
first-hand how our thinking had evolved. Prof.~Dijkstra knew very well
that to effect foundational change, he had to shape the thinking of a
new generation of software professionals and he set about doing so
diligently.

I was fortunate enough to be one of the select few students invited to
join Prof.~Dijkstra's Austin Tuesday Afternoon Club (ATAC). Every
Tuesday afternoon, we would read a technical paper examining the
technical arguments and commenting on how the notations and proofs
could be improved. At one of the ATAC sessions, Prof.~Dijkstra shared
a note from Prof.~Robert Tarjan on Vizing's Theorem, who wrote, ``I
include a proof that is neither clear nor elegant in the hope that you
will rise to the challenge and find the right proof''. I embraced the
opportunity to work on Vizing's Theorem and was able to propose a
simpler solution in the next ATAC meeting. Much to my delight,
Prof.~Dijkstra invited me to collaborate on a manuscript describing
the solution. Not only did this give me an opportunity to work
together with him, but it also gave me insights into the methodical
and disciplined manner in which the master approached his work.

So on February 21, 1990, we began work early, at his study desk in his
home. We worked steadily together, with pen and paper, visiting each
and every detail of the argument, to ensure that the simplest and most
elegant proof was crafted. Prof.~Dijkstra heard me with his
characteristic patience, adopting some of my ideas and suggestions,
while explaining why he discarded others. The argument emerged
iteratively; several preliminary drafts were discarded. After a light
lunch and a walk through their neighborhood together, we resumed
working and by dinner time, a first draft of the manuscript,
\href{https://www.cs.utexas.edu/users/EWD/ewd10xx/EWD1075.PDF}{EWD1075},
was ready. This was revised and refined further twice, mainly by
Prof.~Dijkstra as
\href{https://www.cs.utexas.edu/users/EWD/ewd10xx/EWD1082.PDF}{EWD1082}
and
\href{https://www.cs.utexas.edu/users/EWD/ewd10xx/EWD1082a.PDF}{EWD1082a}.
It gave me a great sense of professional satisfaction to witness
first-hand how one of the great minds in our field worked and I
imbibed some of those practices in my approach to my work as well.

After graduation, I joined IBM Research in 1992. Compared to the
methodical thinking and interactions of my academic upbringing, the
culture of the industrial research environment was very different: it
was replete with fast-talking colleagues whose mannerisms were
influenced by their work and the complexities of the systems that they
were building.

The organization found value in many of the skills that I had acquired
as a graduate student, though not in the ways that I expected. While
my work did not have any impact on the way software is developed at
IBM, practitioners welcomed some of my simple solutions to programming
problems they encountered. My presentations, designed to be crisp and
simple, were appreciated by senior management as they did not
overwhelm them with a morass of detail. Ironically, this made me a
leader in their eyes and despite my reluctance, I was promoted to
technical leadership and management positions. I truly feel that, in
many ways, the success that I have enjoyed at IBM Research, stemmed
from the teachings and values of Prof.~Dijkstra, Prof.~Misra,
Prof.~Apt and others at the University of Texas.

The problems that Prof.~Dijkstra wrote eloquently about, continue to
persist today. While our academic institutions and software industry
have been able to work around these issues with no significant penalty
to the scale of software systems that are built, the problem still
remains. Three decades later, when I see how Programming is taught to
introductory Computer Science students at elite institutions of
learning and I see text-books on Programming, I know that we are
raising a generation who may be adept at today's tools and are quick
to program a solution but lacking a re-examination of their thought
processes may not be able to improve their capabilities
significantly. Given the scale of human ambition, which ever aspires
and spirals higher, it is just a matter of time before we will need
even more advanced techniques. For instance, how are we going to
automate reasoning to support the resurgence of Artificial
Intelligence when we are still in the process of learning how humans
reason?  After three decades in industrial research, I am convinced
that good work never goes waste and that it is only a matter of time
before we will come back to the foundational work that Prof. Dijkstra
began.

I kept in touch with Prof.~Dijkstra and shared my professional
experiences with him. We would communicate by physical mail, since he
didn't use e-mail. He was a faithful correspondent whose beautiful,
multi-page handwritten responses I still have. His letters were a
delightful mix of professional observations, technical issues, life
lessons and witty, personal observations. One of them was completely
written using his left hand. He often urged me to continue the
explorations that we had begun together.  All of them reflected a deep
care and compassion, that I had wrongly mistaken for an aloofness in
our first encounter.

My last discussion with him was in August 2002, just a few days before
he passed away. Sensing that I was distraught, he said, ``When you hear
of my passing, you will be sad and tears will come to your eyes but
then you should let the moment pass and continue with your life for my
life is one that should be celebrated and not be mourned.''

I will always remember Prof. Edsger W.~Dijkstra as an inspirational
teacher and pioneering researcher who had an immeasurably profound
influence on my life. I enclose a photo of me with him from 1989. It
was made in front of the Imperial Abbey of Ottobeuren, Bayern, Germany
during an excursion of the Marktoberdorf Summer School.
\smallskip

\centerline{\includegraphics[width=1\textwidth]{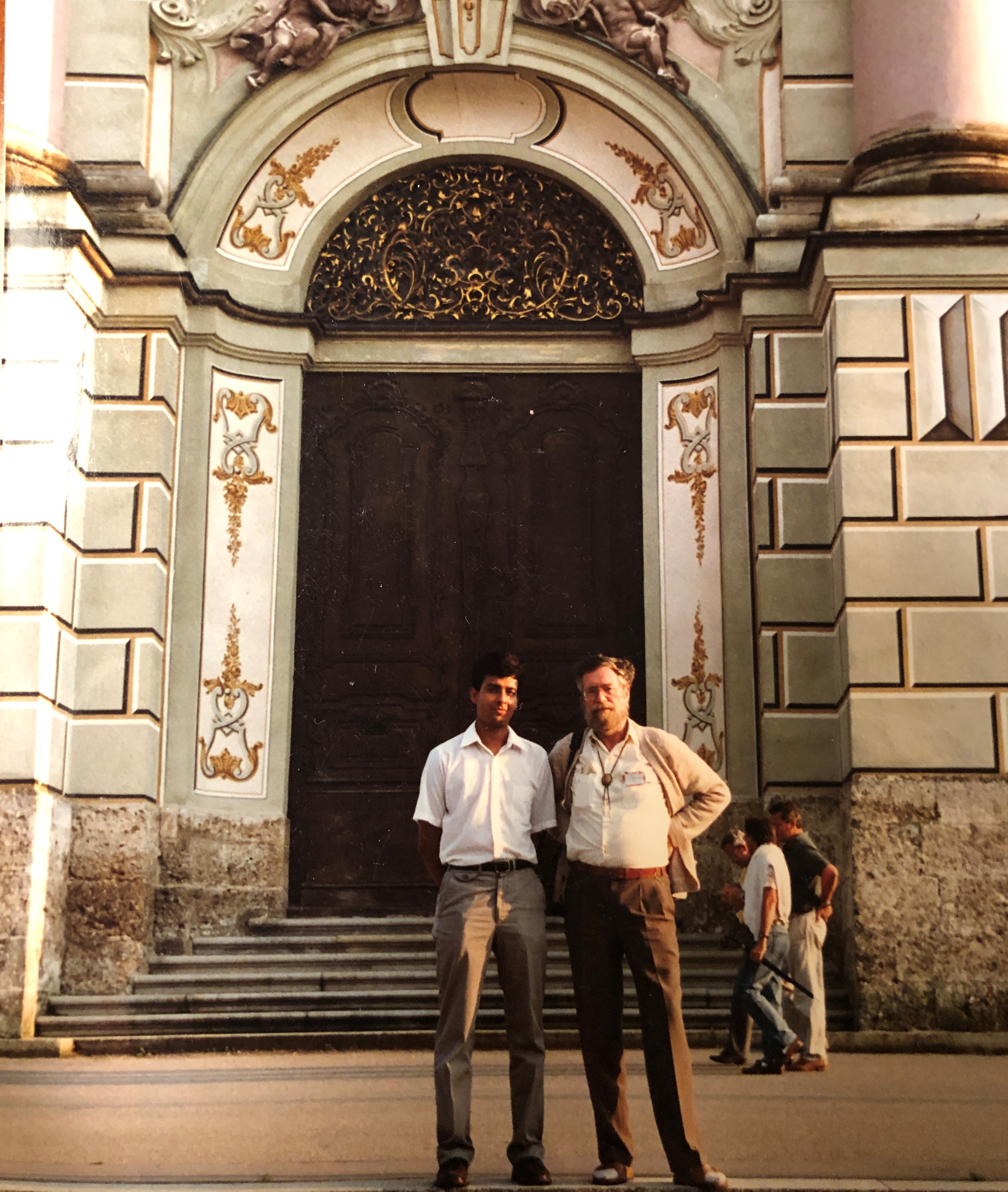}}
\hfill 
\hypertarget{linkB}{\hyperlink{linkA}{$\hookleftarrow$}}

 \newpage

\phantomsection
\addcontentsline{toc}{section}{Jayadev Misra}











\centerline{Jayadev Misra, Computer Science Department, 
University of Texas at Austin}
\centerline{5 March 2021}
\bigskip\noindent
I first met Edsger at a dinner in 1976 when he was visiting Austin as
a Burroughs research fellow. I had a lively conversation with him, and
by the end of the evening he had agreed to mail me his EWD notes. I
met him several times later when he was visiting Austin. I got to know him much
better after he joined the Department of computer science at the
University of Texas at Austin in 1984. I saw him almost every day until
he left Austin in 2002, except the summers which he spent in the
Netherlands. The last time I saw him was in May 2002 at the airport in
Eindhoven; he insisted on seeing me off for a very early-morning
flight because ``he will never see me again''.

I don't remember ever being intimidated by him. I realized that he
will have respect for precise arguments and contempt for shoddy ones,
and he would not be shy about sharing his views in public. So, I was
careful to avoid shoddy arguments in my writing and in speaking to
him. He probably saw potential in me and Mani Chandy. He offered to
read our papers line by line, with us in attendance, and comment on
the writing style, subject matter and, even, on the worthiness of the
paper. Those were exhausting experiences lasting several hours for
even a short paper, but rewarding for they often led to new insights
and crisper arguments.

Dijkstra riled against anthropomorphism, in particular treating
inanimate objects as human beings. He abhorred statements such as
``this guy believes that the other guy is sleeping, so he wakes him up
and the other guy realizes $\cdots$'' to describe communications in a
computer network. So, Mani and I asked him with great trepidation if
he will read our paper ``How processes learn'' sometime during
1985. He actually loved the paper, but asked us to replace ``learn''
by a neutral word. We ignored his suggestion; marketing considerations
prevailed!

About anthropomorphism, here is his sarcastic comment about a
defective toilet: ``It flushes, but without any enthusiasm''.  He told
me about a ticket agent at an airport who said to him ``Mr. Dijkstra,
the computer does not know your name''. I laughed and quipped ``Even
if it did, it would have no respect for you''. His more serious
response was ``That was what attracted me to computers in the first
place''.

Mani, Edsger and I usually went to lunch together where we would
discuss scientific problems. None of us ever noticed what we ate or who
paid the bill. Those are some of the happiest memories of my
professional life. Mani and I were developing the theory of Unity at
that time. We asked Edsger what role the notion of a ``process''
should play in our theory. He drew a matrix on a napkin, explained
that we can partition its elements along the rows, columns or
diagonals. The moral being different views of a system may be
appropriate for different purposes, and a rigid adherence to a static
process structure would stifle a foundational theory. This insight
actually turned out to be a cornerstone in the design of Unity.

He was particularly opposed to the notion of fairness, a concept in
concurrent computing, whereby every participating entity takes a
computational step eventually. He felt, rightly, that the theory of
fairness would be complicated, and, wrongly, that the concept is of
dubious value in practice. He never realized that he was the one who
first introduced the concept in the THE multiprogramming system, a
fact pointed out to me by Amir Pnueli. In designing THE he had assumed
that the peripheral devices operated at finite, but non-zero, speeds.
He was very proud of this abstraction whereby he eliminated the
relative speeds of the devices in his design decisions. He recalled
that his colleagues were horrified that his design did not take into
account the differing speeds of the printer and the punching machine;
he had responded that the punching machine also makes more noise, and
he did not take that aspect into consideration either.

The graduate students in my department used to have a panel discussion
every Friday afternoon. They once invited Edsger and me to debate the
merits of fairness. Edsger went first, delivering a superb argument in
his inimitable style about the unsuitability of waiting for the end of
infinity. I had heard many of those arguments before, so I was
prepared. I claimed that inspired by Dijkstra's talk, I propose
banning irrational numbers, and limiting Turing machine tape length to
the number of atoms in the universe. I believe I thoroughly demolished
his arguments, see \cite{CM-Fairness}.
He shook my hand, formally, after the panel
discussion, but did not say anything. He and his wife, Ria, came to our home
that evening carrying a bouquet of flowers. Dijkstra simply said
``Marvellous response'' and handed me the flowers. 

The \emph{Year of programming} at Austin during 1986-1987 was one of
the highlights of my professional career. It attracted a number of
eminent computer scientists, Tony Hoare, Amir Pnueli and Manfred Broy
among them, who spent extended periods in Austin including a
year-long sabbatical for Tony. There was a series of conferences
attended by many prominent computer scientists from around the
world. The director of the very first conference, Tony Hoare, devoted
the first day of the conference to Unity and asked Mani and me to give
the talks. Unity was then in its nascent stage, so it was a great
opportunity to present the ideas at length, and defend the work
against criticisms thrown from all angles.  Edsger contributed
mightily to it, teasing, probing and suggesting. He attended every
single talk in every conference in that series.

I had once written a very short proof of Cantor's diagonalization
theorem for arbitrary sets using a formal argument, and faxed it to
Edsger. After returning from a month-long trip I received a call from
him in which he announced that ``You are now a co-author with me''. He
told me that he has described the steps for a systematic derivation of
the proof, and that his submitted hand-written manuscript has been
accepted with a few minor revisions, see \cite{cantor00}.
Looking over the paper, I
realized how little of my formal proof was based on a
flash of intuition and how much on a standard sequence of recognizable
steps of which I was completely unaware.

Edsger lived his personal life by the scientific principles he
cherished, one of them being precision in writing and speech. He was
quite fond of my younger son, Anuj, who was extremely precise in his
speech, a habit he had developed on his own from an early age. Once
Anuj was reading a book, \emph{Three men in a boat} by Jerome
K. Jerome, which was way beyond his level of comprehension at the age
of seven. Edsger, who was visiting, looked at the book and asked him
incredulously ``You can read this book?''. Anuj simply said
``Yes'', and after a short pause ``But I can't understand
it''. Edsger's response: ``Touch\'e ''.

At my urging he acquired a cell phone\footnote{He abhorred the
modifier \emph{cell} for a phone because it confused function with
the underlying technology; he much preferred \emph{mobile}
phone.}. He came to my home that evening and asked for help in
entering a password. I found that the user is expected to listen to a
sequence of commands and respond to each of them using the keyboard of
the phone. I managed to get to the point where the phone starts
issuing commands, and then handed it over saying ``Edsger,
listen''. He listened for a while, stopped, looked at me accusingly
and said ``Do you want me to just listen, or listen and act upon it?
Be a computer scientist''.

He enjoyed giving and receiving gifts. My wife, Mamata, and I
discovered a pair of expensive outdoor-chairs left outside our home
sometime in early summer of 1985, with an attached note in Dutch from
which we could only decipher $12\frac{1}{2}$. It was
clearly the Dijkstras who had left the chairs but why, and what is
$12\frac{1}{2}$? We had to wait until their return at the end of the
summer to learn that $12\frac{1}{2}$ years of marriage, our then
marital stage, is celebrated as an anniversary in the Netherlands.

He felt that I am ruining my fine (his words) hand-writing by using
cheap ball point pens; that was the beginning of a series of gifts of
Montblanc pens. And that every educated man needs a true writer's
dictionary, Webster's new universal unabridged dictionary, dangerously
heavy at $2,347$ pages with $320,000$ definitions, which I still
consult. My parents in India enjoyed looking at the miniature bicycle
they had received from the Dijkstras.  Edsger greatly appreciated my
gift of an elegant wooden and glass box to display his collection of
pens. Alas, it could hold only thirty six.

My friendship with him enabled me to enforce rules during my
chairmanship of the department in the early 1990s, rules that he
vehemently opposed. One was to banish all paper memos sent from the chair's
office to the faculty, replacing them by emails. Dijkstra never used a
computer, so, I decreed that the emails be sent to his fax machine at
home. He did not complain; in fact, he may have welcomed it because he
spent a good deal of time at home.

Dijkstra's reputation for searching questions, and sometimes
unreasonable adherence to precision, was legend. Visitors to the
department prepared their talks with great care, yet they were often
found wanting. He questioned not only the technical material in a
talk, but also the style of presentation. A typical question: ``Is
there any coherence in the color scheme on your slides''? Or, to older
visitors who should know better than to recycle a talk: ``What is the
average age of your slides''? I persuaded him to moderate his
questions for younger visitors who were seeking faculty positions, and
hold the questions till the end of the talk rather than cause regular
interruptions. I could tell that Edsger was chafing at this
straightjacket, but what are friends for.

Women had considerable influence in his life starting with his
mother. He often told me how elegantly she solved mathematical
problems, and that he learnt much from her about precise, compact and
beautiful mathematical arguments. I gathered that mother and son were
emotionally very close.  Once when he was in a room with his mother he
thought about a musical composition to play on the piano, but changed
his mind about the piece while walking to the piano. After he
completed playing his mother remarked that she thought he was going to
play the other piece. His wife, Ria, was his closest friend as long as
we knew them. They shared nearly every experience which even included
Ria attending his special talks at the university. In their personal
life she was the leader and he was the obedient follower. His student
Netty van Gasteren lived in the Netherlands but she often spent
extended periods in Austin, staying with the Dijkstras, so that they
could work jointly on scientific monographs. Shortly before his death
Edsger was devastated to learn that Netty was terminally ill; she died
a month after Edsger.

A notable quote (with slight paraphrasing) from \cite{EWD:EWD340pub}:
``In their capacity as a tool, computers will be but a ripple on the
surface of our culture. In their capacity as intellectual challenge,
they are without precedent in the cultural history of
mankind''. Though I am hesitant about accepting the first part of the
claim --I believe as a tool computer will be as fundamental as
electricity-- the second claim, about its influence on our culture,
can be assessed only at a sufficiently long time in the future.  If
this prediction comes true, then Edsger himself will have been the
inspiration for it.  It has been my privilege to learn from a man
whose teaching methods follow the methods of Socrates (elenchus), and
whose reputation is due not just to his writings but to the Dialogues
of his pupil Plato.  That contribution to culture has lasted over two
thousand years.

\hfill 
\hypertarget{linkB}{\hyperlink{linkA}{$\hookleftarrow$}}



 \newpage

\phantomsection
\addcontentsline{toc}{section}{Rajeev Joshi}
\setlength\parindent{0pt}
  
\newcommand{\rotr}{\mathsf{rotr}}
\newcommand{\addp}{{\cal A}}
\newcommand{\fact}[1]{(#1)!}
\newcommand{\spc}{\hspace{2em}}

\smallskip
\centerline{Rajeev Joshi}
\centerline{8 March 2021}
\bigskip
\noindent
I first met Edsger when I took his course, {\em Capita Selecta}, at
the University of Texas at Austin.  The grade was based on a private
oral examination in Edsger's office.  I remember walking into the
office, understandably nervous.  But we started with small chat, and a
cup of coffee (in a mug with the slogan, {\em Rule 0: Don't make a
  mess of it.}, of which I now have a copy).  That settled me, and
minutes later, we had started a journey exploring graphs of functions.
By the end of the exam, we had derived two well-known results from
number theory.  Along the way, Edsger played the role of guide, gently
nudging when necessary, but mostly letting me find the proofs for
myself.  It was easily the most unusual, but also most exhilarating,
exam I had ever taken. Twenty seven years later, I still remember the
derivations, and would like to share them here. \\

We're interested in total functions over finite domains.  Let $S$ be a
finite, nonempty set, and let $f$ be a total function of type
$S \rightarrow S$.  We associate $f$ with a directed graph, where the
nodes are elements of $S$, and $x \rightarrow y$ is an edge in the
graph whenever $f.x = y$.  Since $f$ is total, its graph satisfies\\

(P0) every node has outdegree 1 \\

{\bf Example.} Let $S$ be the set of binary strings of length 3 and
let $\rotr$ be the function that rotates right (with wraparound).  The
graph of $\rotr$ consists of self-loops at the nodes $000$ and $111$,
and the two cycles
$001 \rightarrow 100 \rightarrow 010 \rightarrow 001$ and
$011 \rightarrow 101 \rightarrow 110 \rightarrow 011$. ~~(End of
Example.)\\

We are specifically interested in {\em bijections}, functions that are
one-to-one and onto.  The graph of a bijection satisfies the following
additional properties:\\

(P1) every node has indegree at least 1 \hfill (onto) \\
(P2) every node has indegree at most 1 \hfill (one-to-one) \\

From (P0)--(P2) we conclude:\\

(P3) the graph of a bijection is a collection of disjoint cycles \\

A notable bijection is the identity, whose graph has self-loops at each node.\\

Given the graph of a function $f$, we can compute the graph of $f^2$
by drawing edges between nodes that are at distance two in the graph
of $f$, and similarly for higher powers.  Suppose that in the graph of
$f$ we have a node $x$ belonging to a cycle of length $k$.  In the
graph of $f^k$, node $x$ has a self-loop, and in general, for any $m$,
node $x$ has a self-loop in the graph of $f^{m\cdot k}$.  From this,
we conclude \\

(P4) Let $m$ be a common multiple of the lengths of the cycles in the
graph of a bijection $f$.  Then $f^m$ is the identity. \\

{\bf Example.} The graph of the function $\rotr$ above consists of
cycles of length $1$ and $3$.  Thus $\rotr^3$ is the identity. (End of
Example.) \\

The converse of (P4) gives us \\

(P5) Let $f$ be a bijection such that $f^k$ is the identity for some
positive $k$.  Then every cycle in the graph of $f$ has a length that
divides $k$. \\

A special case arises when $k$ is prime:\\

(P6) Let $f$ be a bijection such that $f^p$ is the identity for some
prime $p$.  Then every cycle in the graph of $f$ has length either $1$
or $p$. \\

The cycles of length $1$ being the fixpoints of $f$, we now have our main result: \\

{\bf Theorem.} Let $f$ be a bijection on a finite domain $S$ such that $f^p$ is the identity for some prime $p$.   Then
\[ |S| - (\#\mbox{fixpoints of } f)   \;\;\;\;\;\; \mbox{is divisible by $p$} \spc \hfill \spc \mbox{(End of Theorem.)}\]

{\bf Corollary 0.} Let $p$ be prime and let $S$ be the set of strings
of length $p$ over an alphabet with $n$ symbols, and let $\rotr$ be
the function that rotates right with wraparound.  We have $|S| = n^p$.
Clearly $\rotr^p$ is the identity.  The only fixpoints of $\rotr$ are
strings with the same symbol repeated; there are $n$ such fixpoints.
Applying the theorem, we conclude
\[ n^p - n \;\;\;\;\; \mbox{is divisible by $p$} \]
which is well known as {\bf Fermat's Little Theorem}. \\

{\bf Corollary 1.} Let $S$ be the set of circular arrangements of the
first $p$ natural numbers.  (Here, ``circular'' means two arrangements
are the same if one can be obtained by rotating the other.) Let
$\addp$ be the function that adds $1$ modulo $p$ to each element in an
arrangement.  Clearly $\addp^p$ is the identity.  Also,
$|S| = \fact{p-1}$ and the fixpoints of $\addp$ are the arrangements
in which each element and its clockwise neighbor differ by $d$ for
some fixed positive value $d$.  Since there are $p-1$ choices for $d$,
the function $\addp$ has $(p-1)$ fixpoints.  Applying the theorem:
\[ \fact{p-1} \; - \; (p-1) \;\;\;\;\; \mbox{is divisible by p} \]
which is known in number theory as {\bf Wilson's Theorem}.\\

I love the simplicity and elegance of the argument, qualities that Edsger embodied so well, and which to this day, continue to guide and inspire.
\hfill
\hypertarget{linkB}{\hyperlink{linkA}{$\hookleftarrow$}}

 \newpage

\phantomsection
\addcontentsline{toc}{section}{Maarten van Emden}


\centerline{\bf An Appraisal}
\centerline{Maarten van Emden, University of Victoria, Canada}
\centerline{27 January 2021}
\bigskip\noindent
I present myself as case study for EWD's pervasive influence.  We had
sporadic encounters between 1968 and 2001.  I started recording in
2008 in Wordpress blogs (\url{https://vanemden.wordpress.com/}),
reaching a total number of around fifty. One of the earliest carried
the title ``I remember Edsger Dijkstra''.  If one searches my
Wordpress site for ``Dijkstra'' almost half of the total number turn
up. The reaction of the visitor must be: ``van Emden has Dijkstra on
the brain''. I can't deny this and that is what I mean by presenting
myself as case study for EWD's pervasive influence.

As a result of a recent change, the
\href{https://www.cs.utexas.edu/~EWD/}{E.W. Dijkstra Archive} has
become a good tribute to EWD.  It would be surprising if any areas are
missing. But consider the ubiquitous eight-bit byte.  Neither this,
nor the very concept of computer architecture, existed when you go
sufficiently far into the past.  EWD played an important role in this
development, see \cite{ve1}.

For obtaining correct code, one can distinguish two levels of
ambition. The first level is to start with code and add invariants and
other assertions later. Finding assertions can be hard; structured
programming alleviates the difficulty.  It was introduced by EWD in
1969 with
\href{https://www.cs.utexas.edu/users/EWD/ewd02xx/EWD249.PDF}{EWD249}
(``Notes on Structured Programming'', April 1970).  The second level
of ambition is to alleviate the difficulty by regarding ``Concern for
Correctness as Guiding Principle for Program Construction''. This is
the title of
\href{https://www.cs.utexas.edu/users/EWD/ewd02xx/EWD288.PDF}{EWD288}
(July 1970) published soon after EWD249. It implicitly
repudiates structured programming.  It is more powerful, if one can
find a way to do it.  Instead Dijkstra went back to structured
programming when he invented guarded commands.  He left EWD288
hanging, leaving as part of his legacy to pick up the challenge, see
\cite{ve2}.

\href{https://www.cs.utexas.edu/users/EWD/ewd02xx/EWD249.PDF}{EWD249}
``Notes on Structured Programming'' is at first sight a search
for a programmer's subliminal thoughts, the processes that occur so
fast that normally they do not penetrate to consciousness.  It is a
search for the atomic building blocks of a programmer's thought
processes.  For example eight pages are devoted to such an analysis of
solving the problem of generating prime numbers.  Then without any
warning on the ninth and final page the treatment becomes extremely
sophisticated.  The end result is an algorithm that was rediscovered
much later.  It takes me many pages to unravel what happens on this
last page, see \cite{ve3}.

For most of his career Dijkstra adhered to the admirable and enviable
discipline of writing down any fruit of his brain that was
write-downable, and to do so as soon as it was write-downable. This
discipline resulted in a long sequence of documents. The E.W. Dijkstra
Archive starts with
\href{https://www.cs.utexas.edu/users/EWD/ewd00xx/EWD28.PDF}{EWD28}
(``Substitution Processes'', 1962) and ends with
\href{https://www.cs.utexas.edu/users/EWD/ewd13xx/EWD1318.PDF}{EWD1318}
(``Coxeter’s Rabbit'', 2002). It is from these that \emph{Selected
  Writings on Computing: A Personal Perspective} (Springer-Verlag,
1982) was compiled.  It contains a report on a trip to attend the IBM
seminar ``Communication and Computers''.  At first sight this is
\href{https://www.cs.utexas.edu/users/EWD/ewd03xx/EWD387.PDF}{EWD387}
(September 1973), but on closer inspection the book version hides the
identities of two speakers by replacing their names by ``NN0'' and
``NN1''.  The former is dismissed in one sentence.  The latter arouses
EWD’s ire so much that he needs a whole page of vituperative prose to
offload his emotions. NN1 is denounced, among other things, for
``appealing to mankind’s lower instincts'' and for ``undisguised
appeal to anti-intellectualism''.  By referring to the original EWD
\href{https://www.cs.utexas.edu/users/EWD/ewd03xx/EWD387.PDF}{EWD387}
NN1 can be identified as the late Douglas Engelbart.

What has Engelbart done to provoke this outburst?  One only has to
refer to ``Engelbart's Law'', see Wikipedia.  Its reasoning seems to
run as follows: look at what mere printing has done as a tool for
thought; the system demonstrated is so much more powerful than
printing that it must quickly lead to Intellect Augmentation.  This
way Engelbart showed no appreciation for the rich culture developed
over centuries.  What makes printing a powerful tool for thought is
mostly due to other things than technology.  Much of the power of this
culture comes from publishers and editors, who sniff out what is worth
printing and hold back what is not.  Another important component of
this culture is provided by libraries and librarians. Much is due to
scholarly societies, which started printing their proceedings and to
commercial publishers, which created journals, each with their
editorial board and unseen bevy of reviewers.  Most of all it is due
to the idea of a university; see for example {\it The House of
  Intellect} by Jacques Barzun (Harper, 1959).  EWD's intemperate
outburst is only a hint at all this.  A further elaboration he left as
part of his legacy, see \cite{ve4}.
Most of this is Engelbart-think;
only at the end it addresses what EWD hinted at in 1973.

An often-heard quote is ``The safest general characterization of the
European philosophical tradition is that it consists of a series of
footnotes to Plato.'' Suitably reduced in scope and scale, this
applies to the legacy of EWD.

\hfill 
\hypertarget{linkB}{\hyperlink{linkA}{$\hookleftarrow$}}



 \newpage

\phantomsection
\addcontentsline{toc}{section}{Two Tuesday Afternoon Clubs}
\centerline{Two Tuesday Afternoon Clubs}

\bigskip

\bigskip

\centerline{\includegraphics[width=1\textwidth]{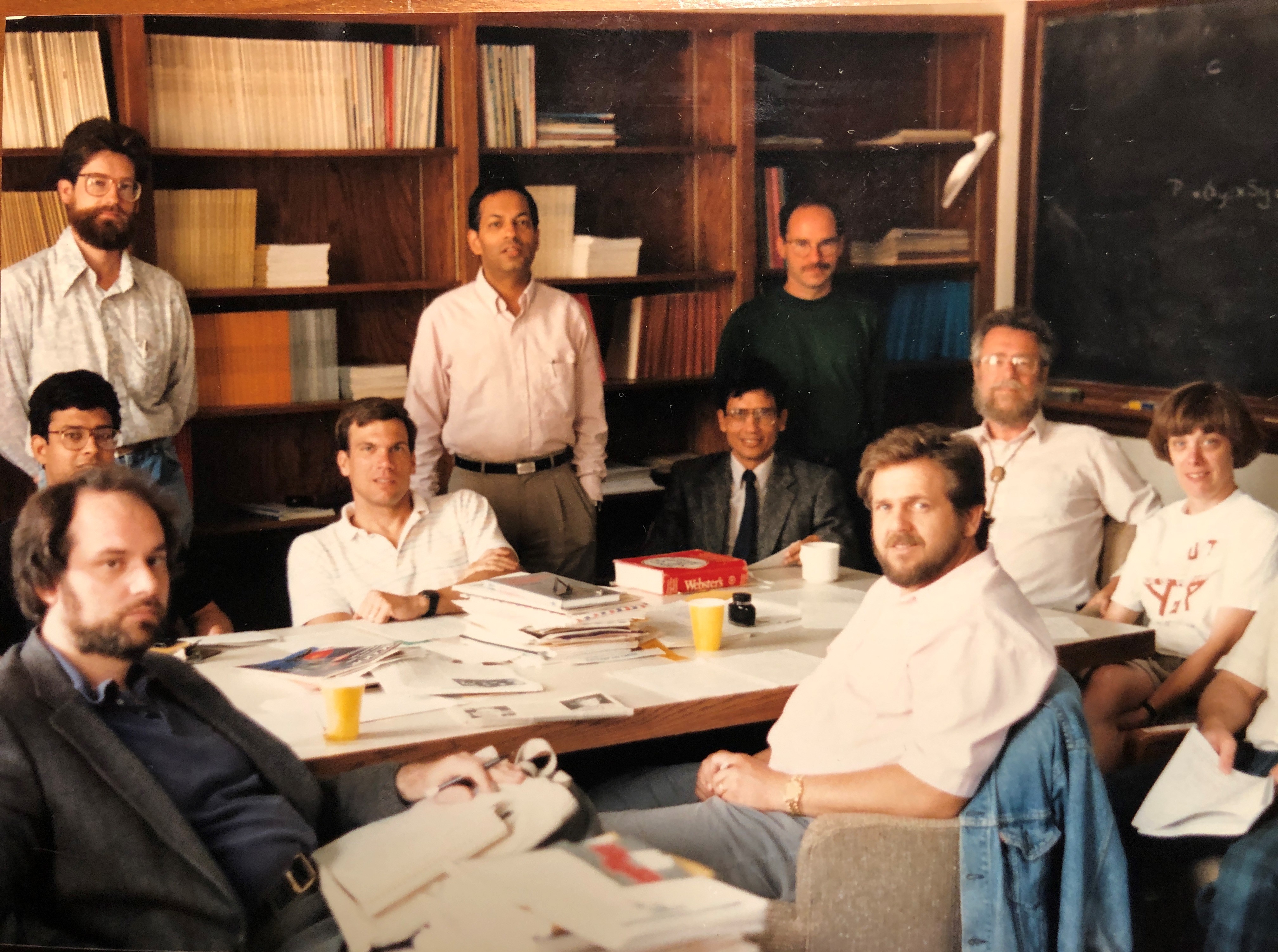}}

\centerline{Austin Tuesday Afternoon Club. Photo made by Charanjit Jutla around summer of 1990}
\medskip

\noindent
Left to right around the table: Allen Emerson, J.R. Rao, David Naumann, Ken Calvert, Jay Misra,
Mohamed Gouda, Ted Herman, Edsger Dijkstra, Netty van Gasteren, and Walter Potter. 

Photo by courtesy of J.R.~Rao. 
\newpage

\centerline{\includegraphics[width=1\textwidth]{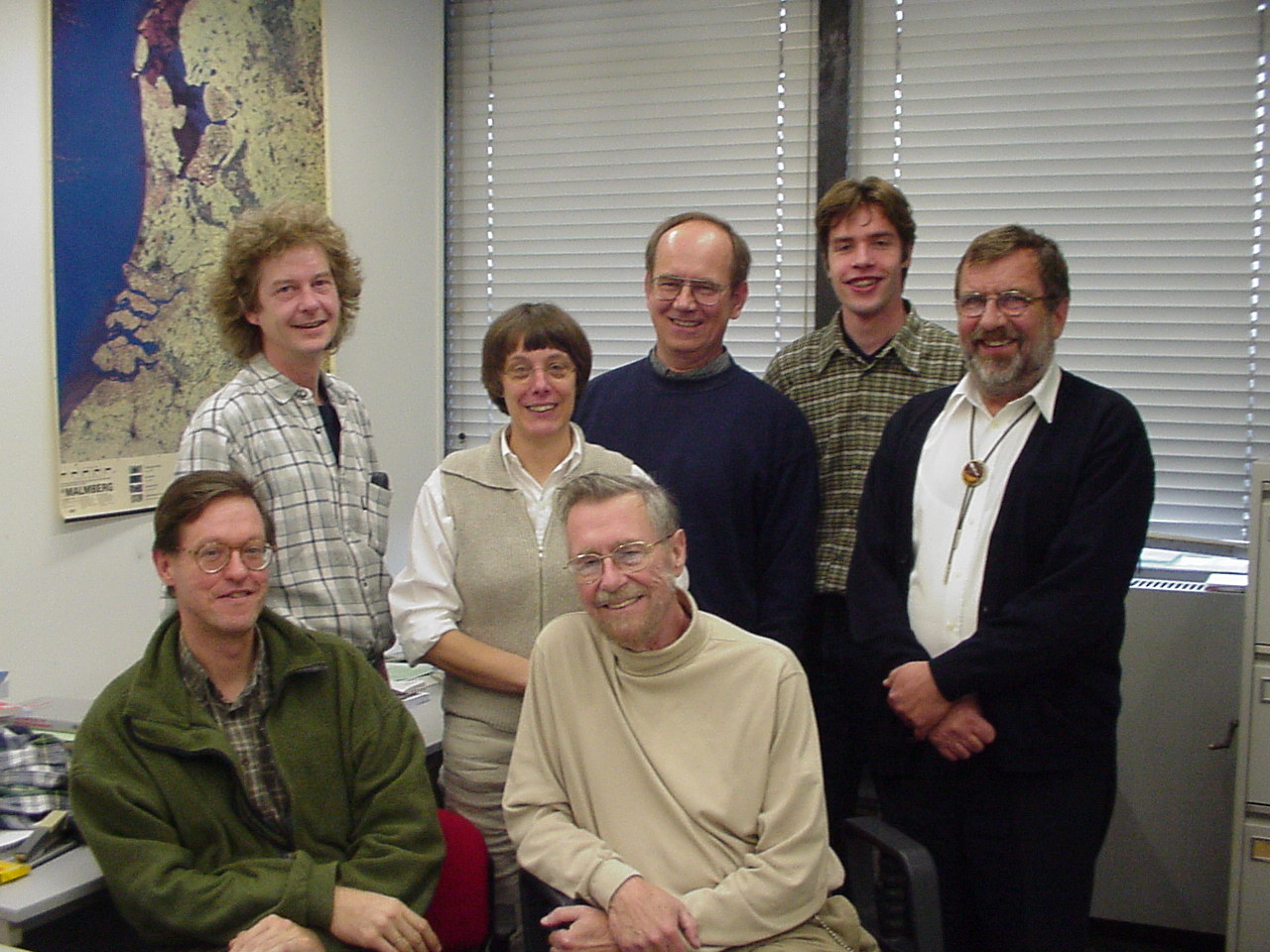}}

\centerline{Eindhoven Tuesday Afternoon Club. The last session during which Edsger was present.}
\centerline{Photo made on Tuesday 18 December 2001}
\medskip

\noindent
Standing from left to right:
Frans van der Sommen,  Netty van Gasteren, Ronald Bulterman, Arjan Mooij and Wim Feijen.
Sitting from left to right:
Gerard Zwaan and Edsger Dijkstra.

Photo by courtesy of Wim Feijen.
\hfill 
\hypertarget{linkB}{\hyperlink{linkA}{$\hookleftarrow$}}

\end{document}